\documentclass[iop]{emulateapj}
\usepackage{url}
\usepackage{hyperref}

\def\correction#1{{#1\/}}
\def\referee#1{{#1}}
\begin{document}
\title{The nonpotentiality of coronae of solar active regions, the
  dynamics of the surface magnetic field, and the potential for large
  flares
%\footnote{First submitted to 
%the Astrophysical Journal ...; resubmitted ...;
%accepted ...0}
%\footnote{\ldots}
}

\author{Carolus J.\ Schrijver}
\affil{Lockheed Martin Advanced Technology Center (A021S, Bldg.\ 252), \\
3251 Hanover Street, Palo Alto, CA 94304 \\
\email{schrijver@lmsal.com}}

\begin{abstract}
  Flares and eruptions from solar active regions are associated with
  atmospheric electrical currents accompanying distortions of the
  coronal field away from a lowest-energy potential state.  In order
  to better understand the origin of these currents and their role in
  M- and X-class flares, I review all active-region observations made
  with {\em SDO/HMI} and {\em SDO/AIA} from 2010/05 through 2014/10
  within $\approx 40^\circ$ from disk center. I select the roughly
  4\%\ of all regions that display a distinctly nonpotential coronal
  configuration in \referee{loops with a length comparable to the
    scale of the active region}, and all that emit GOES X-class
  flares. The data for 41 regions confirm, with a single exception,
  that strong-field, high-gradient polarity inversion lines (SHILs)
  created during emergence of magnetic flux into, and related
  displacement within, pre-existing active regions are associated with
  X-class flares. Obvious nonpotentiality in the \referee{active region-scale
    loops occurs in 6 of 10 selected} regions with X-class
  flares, all with relatively long SHILs along their primary polarity
  inversion line, or with a long internal filament
  there. Nonpotentiality can exist in active regions well past the
  flux-emergence phase, often with reduced or absent flaring. I
  conclude that the dynamics of the flux involved in the compact SHILs
  is of preeminent importance for the large-flare potential of active
  regions within the next day, but that their associated currents may
  not reveal themselves in \referee{active} region-scale
  nonpotentiality. In contrast, active region-scale nonpotentiality,
  which can persist for many days, may inform us about the eruption
  potential other than those from SHILs which is almost never
  associated with X-class flaring.
\end{abstract}


\maketitle

\section{Introduction}\label{sec:introduction}
The ``free energy'' in the corona over solar active regions that
enables flares and eruptions is associated with distortions of the
field away from the potential state, quantified through ${\bf \nabla}
\times {\bf B}$ and reflected in electrical currents that run through
the high atmosphere. \cite{schrijver+etal2005a} compared
potential-field source-surface \citep[PFSS, specifically using the
assimilation-based magnetogram sequence described by][]{schrijver+derosa2002b}
extrapolations over a sample of 95 active regions based on
magnetograms made with {\em SOHO's Michelson Doppler Imager}
\citep[{\em MDI},][]{soho} to extreme ultraviolet (EUV) observations of
their coronae made with the {\em Transition Region and Coronal Explorer}
\citep[{\em TRACE},][]{traceinstrument}. They concluded that ``significant
nonpotentiality of the overall active-region coronal field occurs (1)
when new flux has emerged within or very near a region within the last
$\sim$30\,hr, resulting in complex polarity separation lines, or (2)
when rapidly evolving, opposite-polarity concentrations are in contact
at 4\arcsec\ resolution. If these criteria are met by more than 15\%\
of the region’s flux, they correctly identify the \hbox{(non)}potentiality
of active-region coronae in 88\%\ of the cases.''

The nonpotentiality attending these characteristics of the field
is clearly correlated with flaring: \cite{schrijver+etal2005a}, for example, note
that C-, M-, and X-class ``[f]lares are found to occur 2.4 times more
frequently in active regions with nonpotential coronae than in
near-potential regions, while their average X-ray peak flare
brightness is 3.3 times higher.''  They further argue ``that the
currents associated with coronal nonpotentiality have a characteristic
growth and decay timescale of $\sim$10-30 hr'' and ``that shear flows
drive enhanced flaring or coronal nonpotentiality only if associated
with complex and dynamic flux emergence within the above timescale.''
Yet, nonpotentiality and flaring appear to be related only in a
statistical sense: they find that if ``the occurrence of [C-, M-, and
X-class] flares in the 3 day window is used as a test for potential
and nonpotential coronal fields, only 65\%\ of the regions are
correctly identified [\ldots], which is only 15\%\ better than a
random selection."

The brightest flares on the GOES scale appear to require quite compact
polarity-inversion lines \citep[see the review by][and references
therein]{schrijver2009b}. For example, based on analysis of {\em MDI} and
{\em TRACE} data on 289 M- and X-class flares, \cite{schrijver2007}
suggested that such flaring was generally connected to flux within
some 15\,Mm of ``pronounced high-gradient polarity-separation
lines''. The higher the flux included in such regions, the larger the
maximum flare peak brightness to be expected, and the larger the
probability of such flaring to occur within a given time interval.

Such flare-related strong-field high-gradient inversion lines
(referred to as SHILs below)
appeared to generally accompany flux emergence, as was
confirmed independently by \cite{welsch+li2008}; as flux emergence
into an active region generally makes for intrusions into one or both
polarities of the existing field, flux emergence, flux cancellation,
and displacements involving shearing field are almost unavoidably
going on simultaneously \citep[see, e.g., the review by][with many
references to original studies]{2015RAA....15..145W}.

These two {\em MDI-TRACE} studies referenced above already showed that there
are regions with distinctly nonpotential coronae without exhibiting
SHILs. With the continuous, high-resolution, and full-disk coverage
offered by the Atmospheric Imaging Assembly
\citep[{\em AIA},][]{aiainstrument} and Helioseismic and Magnetic Imager
\citep[{\em HMI},][]{2012SoPh..275..207S} instruments onboard the {\em
  Solar
Dynamics Observatory} \citep[{\em SDO},][]{2012SoPh..275....3P}, I set out
in this study to understand the behavior of the nonpotentiality in
active regions in order to evaluate what role the ``free
energy'' linked with this phenomenon can play in flares and
eruptions.

Active-region scale nonpotentiality could be caused by electrical
current systems attending the flux emergence at SHILs, provided that
the distant field corresponding to these currents is not countered by
essentially oppositely-directed nearby currents of comparable
magnitude.  Magnetohydrodynamic (MHD) simulations provide insight into
the behavior of twisted flux ropes that emerge from within the
convective envelope and currents that are induced by such emergence
and by lateral displacements and twisting motions of photospheric
field.  \cite{manchester+etal2004} and \cite{2012ApJ...754...15F}, for
example, use MHD simulations to show how flux-rope emergence can build
up magnetic shear in the corona through shearing flows at the polarity
inversion line (PIL) and spot rotation.  MHD studies by
\cite{torok+kliem2003}, \cite{2014ApJ...782L..10T}, and
\cite{2015ApJ...810...17D} concur in that non-neutralized, i.e. net,
electrical currents only occur if the twisting and shearing motions of
the flux emergence that they prescribe build up magnetic shear at the
PIL. Thus net currents should be expected to form whenever flows
extend across the PIL or when the stresses they induce involve flux
systems that reach to the PIL.

If the electrical currents linked to, say, M- and X-class flaring and
their coronal mass ejections (CMEs) could be inferred to exist based
on the apparent nonpotentiality of the \referee{most-readily
  observable loops, i.e., those with lengths roughly comparable to the
  scale of the active region \citep[including sigmoid configurations,
  see, e.g., ][and references therein]{2014SoPh..289.3297S}}, then a
powerful tool would become available for the forecasting of flares and
CMEs. Currently, forecasts of flares, and in particular their timing
and magnitude, are of poor quality, with skill scores that are hardly
positive \citep[e.g.][]{leka+barnes2007,barnes+leka2008}. One of the
problems that complicate flare forecasting is that the geometry,
energetics, and stability of the coronal magnetic field cannot be
adequately studied because modeling the field based on surface
vector-magnetic data does not, at present, enable reliable
extrapolations of the field \citep[see][and references therein to a
decade-long series of team studies of nonlinear force-free field
(NLFFF) modeling]{2015arXiv150805455D}. \referee{The principal issues
  appear to be not only the intrinsic nonlinearity of the problem, but
  also the difficulty in inferring all three vector field components
  in an active region and its surrounding quiet Sun, dealing with an
  intrinsic 180$^\circ$ ambiguity in the transverse field component
  that cannot be resolved with single-perspective spectro-polarimetry,
  the unknown forces acting on the observed photospheric field up to
  the top of the chromosphere, and the unknown fields at the sides and
  top of any model volume centered in active regions \citep[problems
  that similarly hamper the application of the magnetic virial
  theorem, as discussed by][]{metcalf+etal2007}.}

Hoping to mitigate \referee{these problems with the use of presently
  available vector-magnetic data for coronal-field studies,
  \cite{2012ApJ...756..153M} and \cite{2015AnA...577A.123C} developed
  methods that can combine surface field observations (line of sight
  and vector fields, respectively) with coronal loop traces to guide
  the NLFFF modeling \citep[see the application to an active region
  by][]{2014ApJ...783..102M}. As these methods use} the set of
active-region loops observable in EUV images (such as by {\em
  SDO}/{\em AIA}) we need to understand whether the seat of power for
energetic flares and eruptions has signatures in these images that are
readily observable and that can be reliably quantified.

In view of these problems, and the above observational and
modeling studies, I use {\em SDO} data in this study to constrain how 
currents involved in flaring distort active-region coronae
away from mostly potential configurations.

\begin{table*}[t]
  \label{tab:summary}\caption{Selected target regions. Column 2 lists
    the short-form SOL 
    identifier specifying time of observation
    (Leibacher et al., 2010), and column 3 the NOAA active region
    number or, if unnumbered, the $(x,y)$ coordinates relative to disk
    center in arcseconds. Column 4 lists the dominant behavior for the surface
    line-of-sight field (capitals) and for the nonpotentiality of the
    loops in comparison with the PFSS field model (lower case
    letters), as detailed in Sect.~\ref{sec:observations}; a minus sign following 
    this indicates absence of sunspots or pores, and an asterisk
    that an eruption preceded
    the observation with ongoing field evolution. 
    Values of $\log(R)$ (in units of $2.2\times
    10^{16}$\,Mx, see Schrijver [2007] for rationale) are listed as
    per the algorithm by Schrijver (2007), based on {\em SDO}/{\em HMI}
    magnetograms scaled to the {\em SOHO/MDI} resolution as used in that
    study; for regions with X-class flares, the ratio of
    $R$ to total absolute flux $\Phi$ is also listed. Column 6 lists flares above M5,
    if any; the GOES flare class is followed by ``N'' if the flare was
    non-eruptive as inferred from {\em SDO}/{\em AIA} and SOHO/LASCO
    observations.}
\begin{center}
\begin{tabular}{rccrccl}
\hline\hline 
No. & SOL & AR no. & Cat. & $\log(R)$,& Flare(s) & Notes \\
       &       &             &        & $\log(R/\Phi)$ & & \\ 
\hline 
1 & 2010-07-01T22:29:00 & 11084 & Eac & 2.5 & - & Repeated ``filament'' act.\ from AR to QS to W, SW\\
   & 2010-07-02T22:29:00 &    ``    & Eac & 0.0 & - & Repeated ``filament'' act.\ from AR to QS to W, SW\\
2 & 2010-08-02T22:29:00 & 11092 & Fab & 2.6 & - & Erupt's to SW and NE. Cf.\ Schrijver {\&} Title (2011)\\
   & 2010-08-03T22:29:00 &    ``    & Fab & 0.0 & - & Erupt's to SW side\\
   & 2010-08-04T22:29:00 &     ``   & Fab & 2.1 & - & Erupt's to SW and NE side\\
3 & 2011-02-13T16:00:00 & 11158 & ABc(?)e & 4.6 & M6.6 & - \\
   & 2011-02-15T00:26:00 &    ``    & Bc(?)e & 4.6, -1.4 & X2.2 &  Cf.\ Schrijver et
al. (2011)\\
4 & 2011-02-23T22:29:00 & $(-64,-95)$    & Eab- & 0.0 & - & - \\
5 & 2011-03-09T21:53:00 & 11166 & ABc & 4.7 & X1.5, -1.5/N& Several large-scale
fronts from sides\\
6 & 2011-03-27T22:29:00 & 11176 & Abe & 3.7 & - & Large eruption NW
extremity $\approx$0\,UT\\
   & 2011-03-28T22:29:00 &    ``   & Aabe & 3.3 & - & Mild activity
   in interior and towards NE\\
   & 2011-03-29T22:29:00 &    ``   & Abe & 3.3 & - & -\\
7 & 2011-06-19T22:29:00 & 11236 & Ed & 2.4 & - & Conf.\ erupt's
$\sim$2:30, $\sim$12:30, $\sim$16:30, $\sim$18:30\,UT\\
   & 2011-06-20T22:29:00 &    ``    & Ed & 2.9 & - & Conf.\ erupt's
$\sim$7:00, $\sim$18:30, $\sim$22:00, $\sim$23:30\,UT\\
8 & 2011-06-23T22:29:00 & W of 11240 & Cac- & 0.0 & - & No notable impulsive
events on 2011/06/23\\
9 & 2011-09-06T20:50:00 & 11283 & Bc & 4.1, -2.0 & X2.1 & - \\
   & 2011-09-07T21:08:00 &    ``   &  Bcd & 4.0, -2.0 & X1.8 & Associated with eruption of cool material\\
10 & 2011-09-13T22:20:00 & 11289/-93 & E(C)cde* & 3.8 & - & Large eruption and post-erupt.\ loops at end of day\\
11 & 2011-10-27T22:29:00 & 11330 & Cae & 4.0 & - & Large QS eruption on the N to NW side $\approx$midday\\
     & 2011-10-28T22:29:00 &   ``     & Cade & 4.1 & - & Eruption on NE side $\approx$16\,UT\\
12 & 2011-12-02T22:29:00 & 11362 & ACe & 3.5 & - & No notable impulsive
events on 2011/12/02\\
     & 2011-12-03T22:29:00 &   ``     & ACe & 3.5 & - & Internal act.\ $\approx$05\,UT and in N-periphery $\approx$14\,UT\\
     & 2011-12-04T22:29:00 &    ``    & Ce & 2.8 & - & Internal act.\ $\approx$16\,UT\\
13 & 2012-01-18T22:29:00 & 11399 & Ebc* & 0.0 & - & During afterglow of a long AR-QS filament eruption\\
14 & 2012-03-06T22:54:00 & 11429/-30 & BCbce & 5.0, -1.2 & X5.4/X1.3 & Frequent activity within -29 and in loops to -30\\
     & 2012-03-08T22:29:00 &    ``  & BCbce & 4.7 & - & Quiescent (but with data gaps)\\
     & 2012-03-09T22:29:00 &    ``  & Bbce & 4.6 & - & Major eruption $\approx 3$\,UT\\
15 & 2012-03-28T22:29:00 & 11442 & Cbce & 3.4 & - & Recently emerged next to existing bipolar region\\
16 & 2012-05-03T22:29:00 & 11470/-1/-2 & (C)abe & 3.3 & - & Moderate mostly confined eruption $\approx$11:30\,UT\\
     & 2012-05-04T22:29:00 &     ``   & Eabe & 0.0 & - & Possible eruption $\approx$15:30\,UT\\
     & 2012-05-05T22:29:00 &    `` & Ebe & 3.3 & - & Large eruption $\approx 06$\,UT\\
17 & 2012-05-28T22:29:00 & 11490 & Ce & 3.9 & - & Fast QS coronal evolution on leading side $\approx 11$\,UT\\
     & 2012-05-29T22:29:00 &    ``    & C(a)e & 3.5 & - & Fast QS coronal evolution in leading region $\approx$11\,UT\\
18 & 2012-06-04T22:29:00 & 11497 & Dae & 2.1 & - & [Missing {\em AIA} daily movies] \\
19 & 2012-07-10T04:01:00 & 11519/-20/-1 & BCbc & 4.8 & M5.7 & Also frequent moderate activity\\
     & 2012-07-11T22:29:00 &     `` & Bbe & 4.3 & - & Frequent moderate activity\\
     & 2012-07-12T15:19:00 &     `` & Bbe & 4.8, -2.1 & X1.4 & Also frequent moderate activity\\
20 & 2012-07-31T22:29:00 & 11532/-36 & Cabe & 3.9 & - & -\\
     & 2012-08-01T22:29:00 &    ``  & Dabe & 3.6 & - & -\\
     & 2012-08-02T22:29:00 &    ``  & Gabe & 3.4 & - & -\\
\hline
\end{tabular}
\end{center}

\end{table*}
\nocite{schrijver+title2010}\nocite{2011ApJ...738..167S}\nocite{2010SoPh..263....1.}

\addtocounter{table}{-1}
\begin{table*}[t]
\caption{(cnt'd)}
\begin{center}
\begin{tabular}{rccrccl}
\hline\hline 
No. & SOL & AR no. & Cat. & $\log(R)$, & Flare & Notes \\
       &       &             &        & $\log(R/\Phi)$ & & \\ 
\hline 
21 & 2012-08-11T22:29:00 & 11542 & Fa* & 3.9 & - & Eruption $\approx$16:30\,UT\\
22 & 2012-09-01T22:29:00 & 11555/-60/-61 & Cbce* & 4.3 & - & Frequent eruption from region and nearby\\
23 & 2012-09-17T22:29:00 & 11569/-71/-74 & E(C?)ae & 2.9 & - & - \\
24 & 2012-10-28T22:29:00 & 11599 & -a & 0.0 & - & Isolated spot, no sign.\ coronal activity\\
25 & 2012-11-29T22:29:00 & 11621 & EGde & 2.3 & - & Only limited activity\\
     & 2012-11-30T22:29:00 &    ``    & EGde & 2.3 & - & Only limited activity\\
     & 2012-12-01T22:29:00 &    ``    & EGde & 1.8 & - & Only limited activity\\
26 & 2012-12-02T22:29:00 & 11623/-25 & Cabe & 2.0 & - & No substantial activity\\
     & 2012-12-03T22:29:00 &    `` & Ebe & 3.2 & - & No substantial activity\\
27 & 2012-12-19T22:29:00 & 11633/-4 & Ga & 3.6 & - & Moderate erupt's
AR\,11633 $\approx$02, 10, 18, {\&} 24\,UT\\
28 & 2013-05-17T22:29:00 & 11745 & Fce & 2.6 & - & -\\
     & 2013-05-18T22:29:00 &     `` & Fce & 2.3 & - & -\\
29 & 2013-06-10T22:29:00 & $(163,38)$ & Eabe- & 1.5 & - & -\\
     & 2013-06-11T22:29:00 &    `` & Eabe- & 0.0 & - & -\\
30 & 2013-10-13T22:29:00 & 11864/-5 & Abc* & 4.2 & - & Eruption $\approx$00:30\,UT and $\approx$20\,UT\\
     & 2013-10-14T22:29:00 &     `` & Cbc & 4.1 & - & Erupt.\ $\approx$13\,UT, and 
from leading flux $\approx$16:30\,UT\\
31 & 2013-11-08T02:56:00 & 11890 & Cc & 4.5, -1.9 & X1.1 & Compact intrusion at trailing polarity\\
     & 2013-11-10T03:44:00 &    ``   & Cd & 4.3, -1.9 & X1.1 & Compact intrusion at trailing polarity\\
32 & 2013-11-14T22:29:00 & 11895/-7 & Ccde & 4.3 & - & - \\
     & 2013-11-15T22:29:00 &     `` & Ccde & 4.1 & - & - \\
     & 2013-11-16T22:29:00 &    ``  & Ccde & 4.0 & - & Eruption $\approx$08\,UT\\
     & 2013-11-17T22:29:00 &    ``  & Ccde & 3.8 & - & Repeated
     jet-like activity\\
     & 2013-11-18T22:29:00 &    ``  & Ede & 3.0 & - & Some moderate activity\\
33 & 2014-01-07T17:02:00 & 11944 & Cae & 4.7, -1.8 & X1.2 & X-flare not over
high-gradient field. Conn.\ to 11946\\
34 & 2014-01-08T22:29:00 & 11946 & Aae & 3.4 & - & Moder.\ internal act.\
and northward. Conn.\ to 11944\\
35 & 2014-01-13T22:29:00 & 11950 & Ea & 2.8 & - & QS filament eruption
trailing AR\\
36 & 2014-02-06T22:29:00 & 11970 & G(C?)bce* & 2.7 & - & Signs of
eruption $\approx$21\,UT\\
37 & 2014-03-29T17:48:00 & 12017 & ABc            &  4.3, -1.7 & X1.0 &
Preceded by at least two flares/eruptions\\
38 & 2014-05-25T22:29:00 & 12071/-3 & Cb & 2.9 & - & -\\
     & 2014-05-26T22:29:00 &     `` & Cc & 2.9 & - & -\\
39 & 2014-06-16T22:29:00 & 12090 & Ea & 2.7 & - & -\\
     & 2014-06-17T22:29:00 &     `` & Ea & 2.2 & - & -\\
40 & 2014-09-10T16:15:00 & 12158 & Gcd & 4.3, -1.8 & X1.6 & Moderate
activity in and around AR\\
41 & 2014-10-22T12:58:00 & 12192 & Ad & 5.1, -1.5 & X1.6/N & Very active;
see text\\
     & 2014-10-23T08:20:00 &     ``   & Ad & 5.0 & M1.1/N & Very active;
see text\\
     & 2014-10-24T20:16:00 &     ``   & Ad & 5.1, -1.7 & X3.1/N & Very active;
see text\\
     & 2014-10-25T15:38:00 &     ``   & Cd & 5.1, -1.6 & X1.1/N & Very active;
see text\\
     & 2014-10-26T09:26:00 &    ``    & Bd & 5.2, -1.5& X2.0/N & Very active;
see text\\
\hline
\end{tabular}
\end{center}

\end{table*}

\section{Observations}\label{sec:observations}
For the purpose of this study, I selected solar regions observed by
{\em SDO}/{\em AIA} from 2010/05 through 2014/10 based on a subjective criterion:
all regions within approximately 40$^\circ$ from disk center that
appeared substantially nonpotential when comparing a field model with
observed loops in the {\em AIA} 171\,\AA\ channel (generally consistent with
the 193\,\AA\ channel data, not used further in this study). For the
latter assessment, I reviewed the daily overlays of PFSS model field
lines shown on the {\em AIA} ``Sun In Time'' web
pages\footnote{\url{http://sdowww.lmsal.com/suntoday_v2}}; regions
were selected on multiple dates if the criteria of distance to disk
center and appearance of nonpotentiality were met. Most regions have
NOAA active region numbers, but two (one small, one aged) do not.  After
this initial selection based on 1645 daily {\em AIA}-PFSS overlays, I added
any regions within 40$^\circ$ from disk center emitting an X-class
flare if not already included.

For the selected regions, comparisons of {\em AIA} 171\,\AA\ data and PFSS
models were made 1.5\,h prior to flares above GOES class M5 in the 
regions or --~if no such flares occurred~-- 1.5\,h prior to the start
of a UT date without such flares. The set of 41 selected regions with
78 distinct dates is listed in Table~\ref{tab:summary}.

For each of the entries in Table~\ref{tab:summary}
a 4-day sequence of magnetograms was made, remapped to disk center after
removing the differential rotation characteristic of the mean
latitude of the target region. Column~4 in Table~\ref{tab:summary}
lists characteristics of the surface magnetic field (in capital
letters, sometimes in two or more categories, and between parentheses
if weakly or ambiguously), and in the coronal loop configuration in
comparison to the PFSS model field (in lower case letters).

For the characterization of the coronal configuration in comparison to
the PFSS model field, I use the same qualitative criteria as identified
by \cite{schrijver+etal2005a}. That study characterized the
differentiating properties of evolving surface magnetic field for the
sets of nonpotential and mostly potential regions. Quoting directly from
that study (including percentages that were given for the fraction of
regions with such characteristics, sometimes in multiple categories):
\def\customitemize{
  \itemsep=0pt \parsep=0pt \parskip=0pt \partopsep=0pt \topsep=0pt}
\begin{enumerate}\customitemize
\item[A] ({71\%} of all nonpotential cases): field is still emerging,
  or was within the last day, with meandering or fragmented polarity
  separation lines, associated with sustained shearing motions of the
  field as it sorts itself out; in some cases this leads to
  nonpotential fields low down while leaving the largest scales nearly
  potential;
\item[B] ({43\%}): there are touching and rapidly evolving (canceling)
  opposite polarity concentrations of high flux density up to $\sim
  30$\,h earlier, unless that is a small part of the overall flux in the
  region; if these form a relatively small and compact fraction of the
  overall active region, the nonpotentiality is weak or limited to that
  compact region;
\item[C] ({14\%}): field emerges within or adjacent to an
  existing configuration, provided it introduces significant new flux,
  and does not have a neutral line shared with that of the pre-existing
  field.
\end{enumerate}
They also note that the \hbox{(near-)}potentiality
of the overlying field is not (necessarily) affected if (quoting again):
\begin{enumerate}\customitemize
\item[D] ({41\%} of all near-potential cases): 
 strong mixed-polarity emergence has strongly decreased in 
 the last $10-20$\,h, or when such emergence is just starting;
\item[E] (36\%):
 the configuration is that of a simple, gradually-evolving bipole;
\item[F] (18\%):
 flux cancels rapidly in a mature or decaying region
 having a simple bipolar configuration with a relatively straight and
 well-defined polarity inversion  line;
\item[G] (5\%):
 relatively small bipoles emerge within an existing active region,
 regardless of their position relative to the neutral line. 
\end{enumerate}

For the present study, the deviation of {\em AIA} coronal loops from the PFSS
model is characterized in five classes that are  complementary only to a limited degree:
\begin{enumerate}\customitemize
\item[a)] in a large-scale swirl in loops from one or more major flux concentrations;
\item[b)] in large-scale loops system essentially connecting the two main
  polarity regions:
\item[c)] in low-lying coronal emission structures or chromospheric
  absorption structures;
\item[d)] in a small set of loops at a relatively small segment of the
  AR corona;
\item[e)] in loops connecting to surrounding active regions or  distant
  quiet Sun.
\end{enumerate}

Table~\ref{tab:summary} also lists the total flux in the vicinity of
the strong-field, high-gradient polarity inversion lines (denoted
below as SHILs) associated with the active region (or active region
complex if in close proximity) as introduced by \cite{schrijver2007}
(giving the value of $\log(R)$ in the fifth column of the table), and
the largest flare on a given date provided of GOES class M5 or
larger\footnote{Supporting electronic materials are available at
\url{http://www.lmsal.com/nonpotential/}, including links to
the {\em SDO} SunToday pages, remapped magnetogram sequences, $\log(R(t)$
diagrams, the PFSS field line overlays on magnetograms and {\em AIA}
171\,\AA\ images, and {\em AIA} daily summaries.}.

Many of the selected regions, in particular those exhibiting X-class
flares, have been analyzed in the literature. The subsections below
review that work\footnote{Literature studies were identified using
  iSolSearch [\url{http://lmsal.com/isolsearch}] which can perform ADS
  queries for papers with AR numbers in title or abstract.}  where
pertinent to this study, adding some comments to complement or clarify
entries in Table~\ref{tab:summary} and Figs.~1-6.

\begin{figure*}
\begin{center}
\noindent\includegraphics[width=14.7cm]{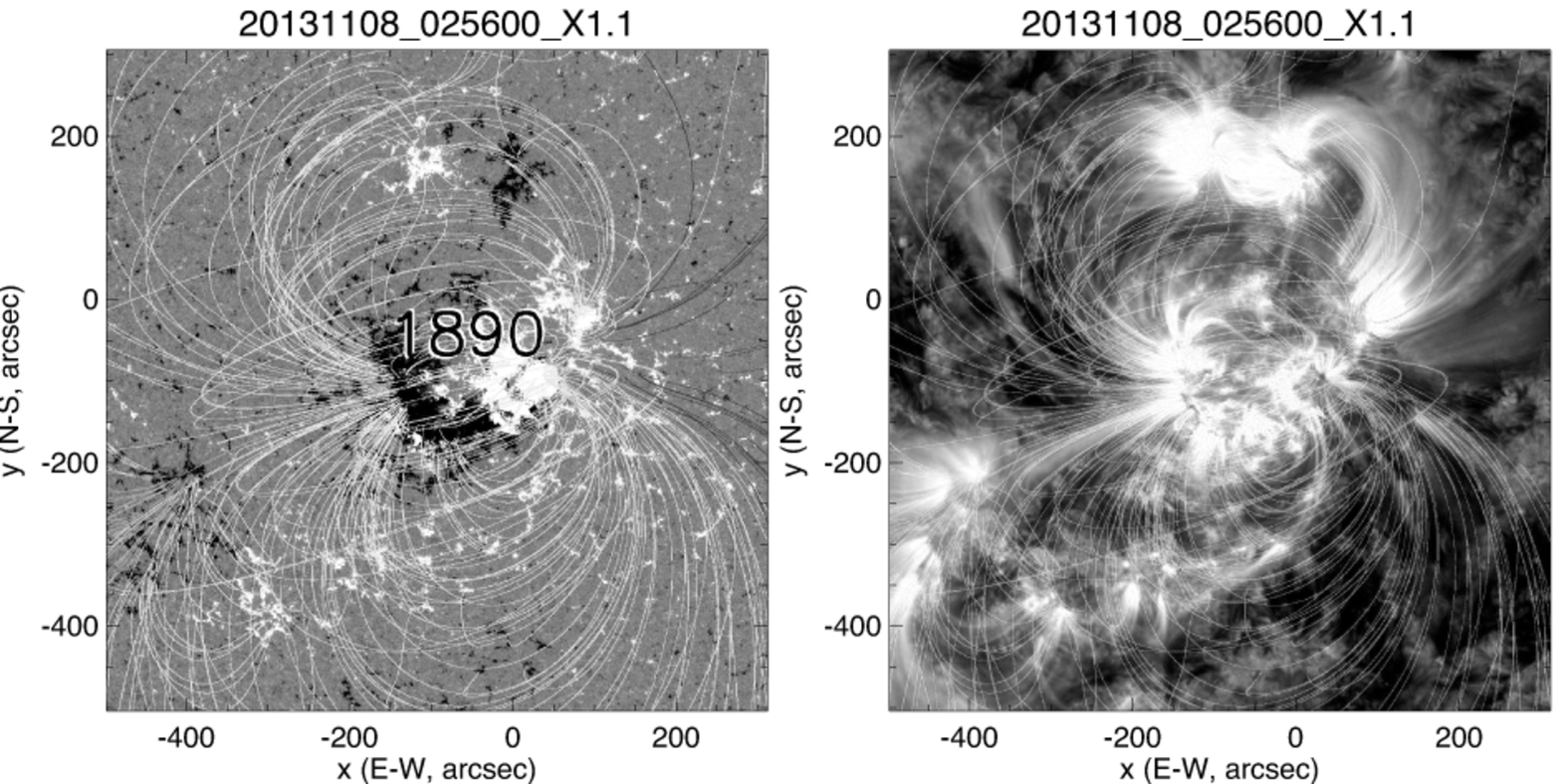} 
\noindent\includegraphics[width=14.7cm]{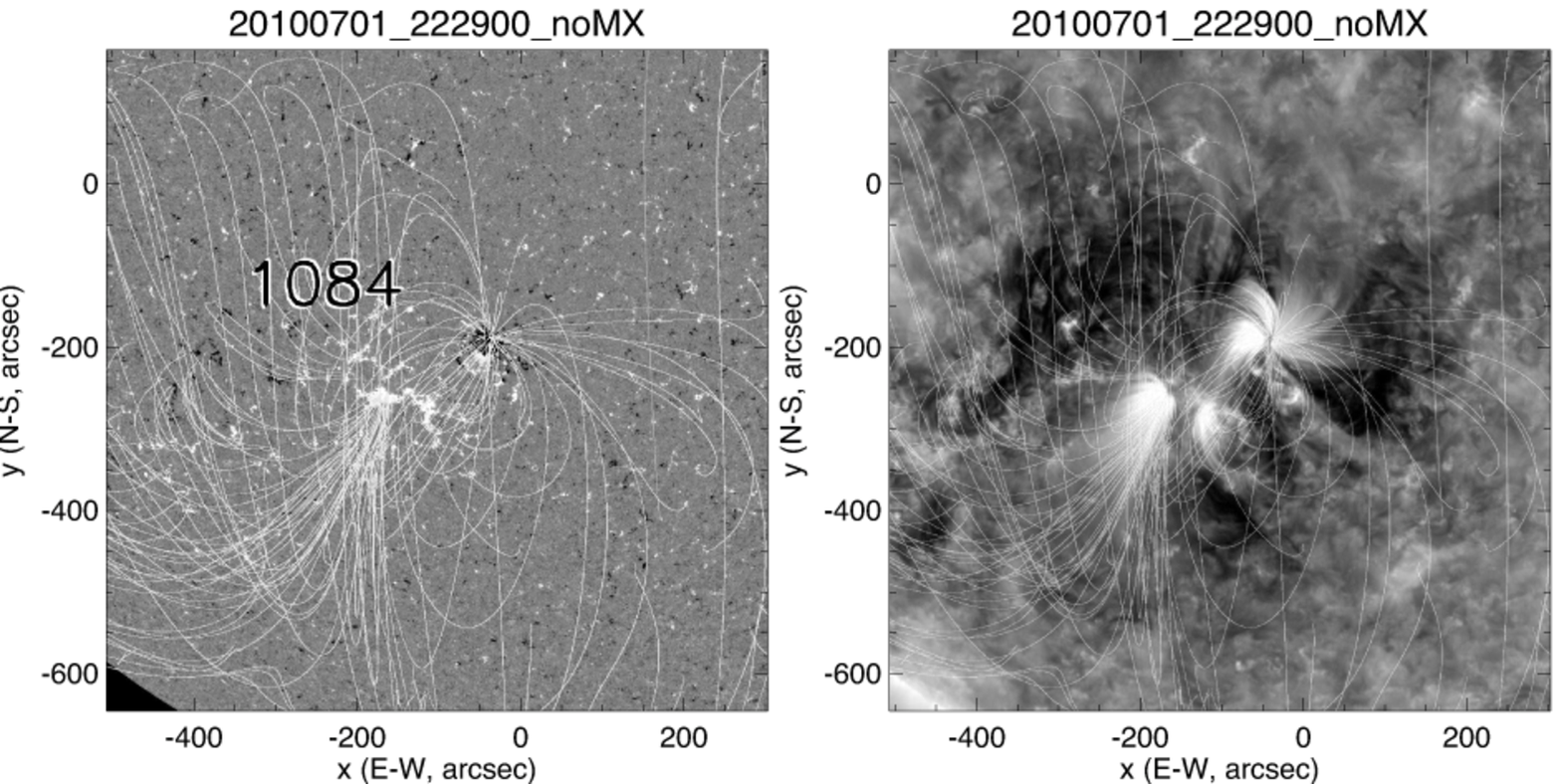} 
\end{center}
\caption{Example regions in category 'c', with and without X flares:
  {\em (top)\,} AR\,11890 {\em (bottom)\,} AR\,11084. Each panel shows
  a field of view of 810\,arcsec (1350 image pixels) to a side. Left
  panels: {\em HMI} magnetograms \referee{(with the positive and
    negative polarities -~shown in white and black, respectively~-
    saturating at $\pm 75$\,Mx/cm$^2$)} with PFSS field lines overlaid
  (white if closed, dark if ``open'' to the heliosphere, i.e. if
  connecting to the source surface); the probability of starting field
  lines at a give location increases with increasing flux
  density. \referee{Only the last four digits of the AR number are
    shown, following the standard practice at NOAA's Space Weather
    Prediction Center.} Right panels: {\em AIA} 171\,\AA\ images with
  the same field lines blended in.  }\label{fig:3}
\end{figure*}
\subsection{AR\,11084} 
The nonpotentiality of the corona over AR\,11084 (shown in the bottom
panels of Fig.~\ref{fig:3}) was also shown by
\cite{2014SoPh..289..475K}, who noted that it persisted for the entire
disk passage. This region has well-separated polarities, exhibiting
nonpotential signatures most clearly in the counterclockwise spiral
in the loops emanating from the leading spot. Although the region does
not exhibit M- or X-class flaring (not expected for a region of this
flux content), it does show repeated activity within it in connections
to the quiet Sun on the west and southwest.

\cite{2013SoPh..287..279L} found that the flows around AR\,11084 at
the surface and at a rather shallow depth of 500\,km were quite
similar, although somewhat different in horizontal extent.
\cite{2014SoPh..289..475K} used ring-diagram analysis to find a
predominant kinetic helicity in the subsurface flows, but the
connection with the coronal nonpotentiality remains unclear, because
two of their six control regions without ``persistent whirl patterns''
in chromospheric and coronal images also showed kinetic helicity of
one predominant sign. 

\subsection{AR\,11092} 
Activity in and around AR\,11092 was described for 2010/08/01 and 02
by \cite{schrijver+title2010}, who focused on the long-range couplings
between the fields involved, including likely influences from emerging
flux on the far hemisphere on the flaring of the region and the
destabilization of neighboring quiet-Sun filaments.

\cite{2013SoPh..287..279L} map the flows around AR\,11092 in
photospheric and subphotospheric layers to find similar flows at
different depths around the main sunspots. The flow patterns around
the area of flux emergence, in contrast, are quite
different: the characteristic surface features of separation of
polarities, rotation of the sunspots, and shear along the
polarity-inversion line do not appear to have counterparts at depth.

As for AR\,11084, \cite{2014SoPh..289..475K} noted a
persistent nonpotential swirl in the field over 
the leading spot of AR\,11092, and also here find a predominant
kinetic helicity in the subsurface flows.

\begin{figure*}
\begin{center}
\noindent\includegraphics[width=14.7cm]{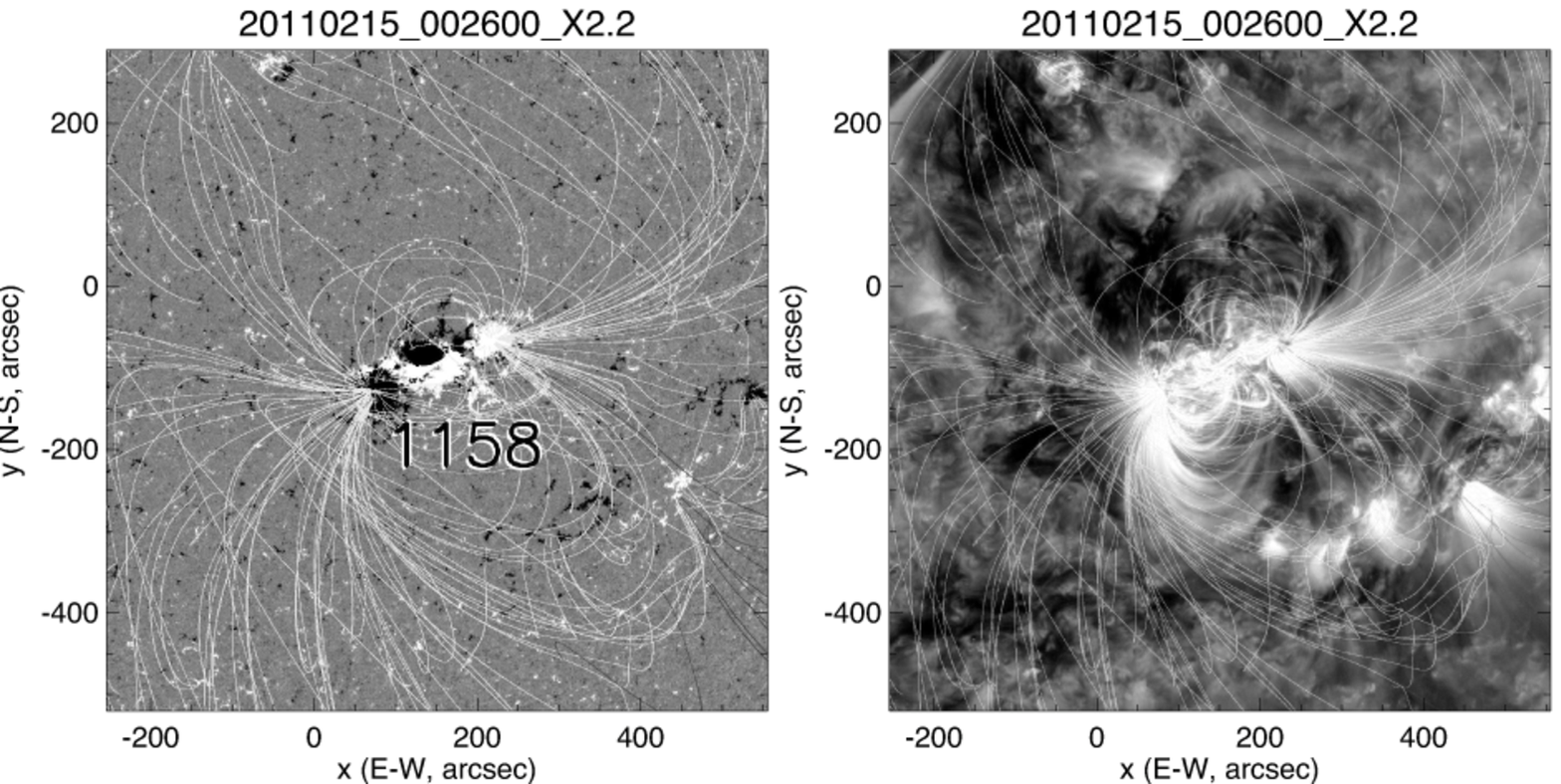} 
\noindent\includegraphics[width=14.7cm]{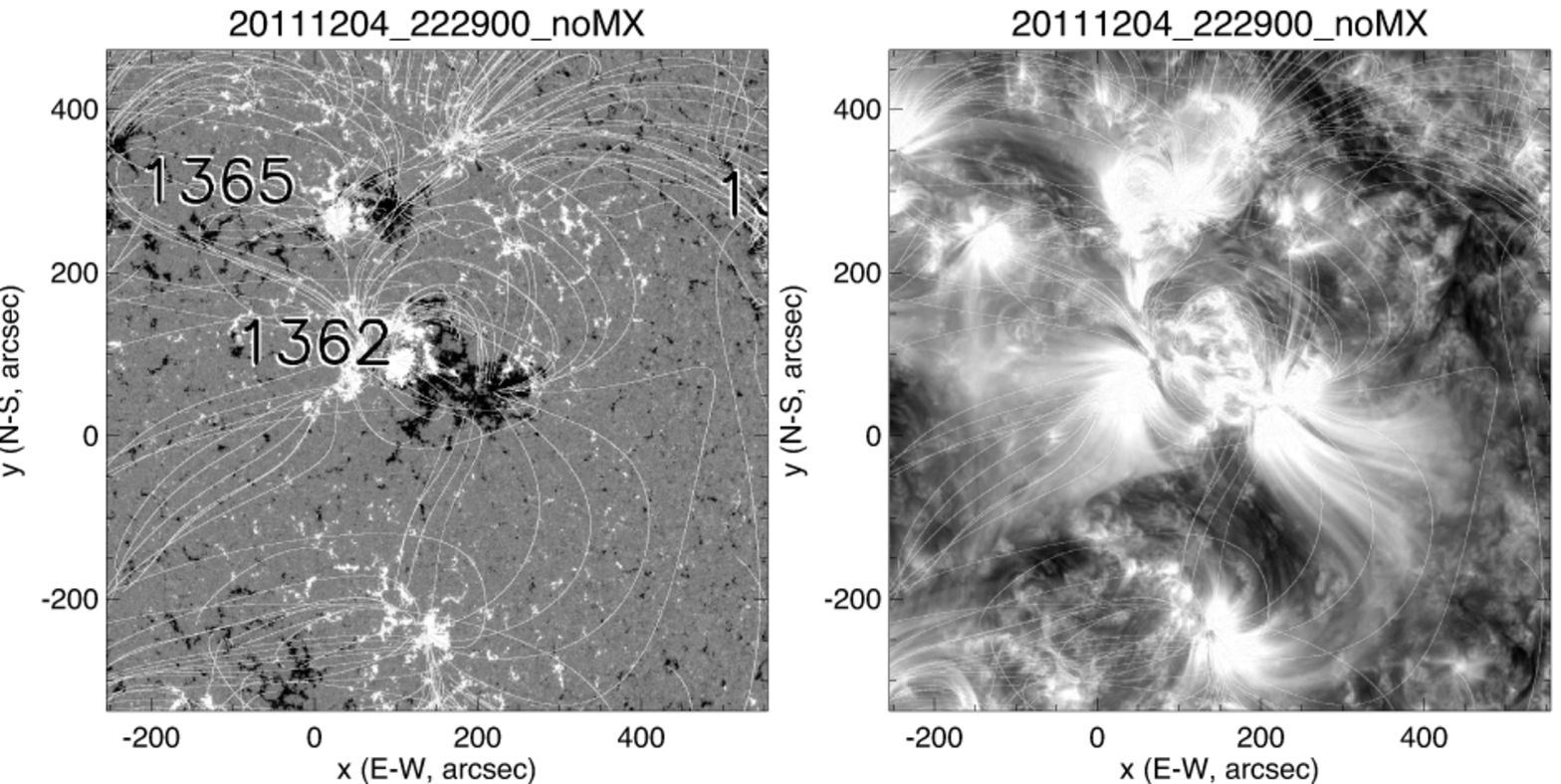} 
\end{center}
\caption{As Fig.~\ref{fig:3}: Example regions in category 'e', with and without X flares:
{\em (top)\,} AR\,11158
{\em (bottom)\,}  AR\,11362.
}\label{fig:5}
\end{figure*}
\subsection{AR\,11158} 
The region (Fig.~\ref{fig:5}, top), has
signatures of nonpotentiality mostly in its interior and in a few
long loops connecting outward from the interior of the
region. Overall, the compatibility of {\em AIA} 171\,\AA\ loops and PFSS
field lines is quite difficult to assess. Loop bundles on the southern
side, for example, are not outlined by field lines, but this could be
a sampling problem. Clearly, there is no dominant swirl in loops from
either of the dominant flux clusters at the region's ends.

AR\,11158 produced the first X-class flare (X2.2) of the current
sunspot cycle, a flare that was the first of its class observed by
{\em SDO}/{\em AIA}. This event consequently received much attention in the
literature, starting with
\cite{2011ApJ...738..167S}. \cite{2014SoPh..289.3351T} \citep[see
also][]{2014ApJ...788...60J} describe the evolution of the surface
field of the region and of its coronal configuration up to the point
of the X2.2 flare. They apply MHD modeling of the flux emergence to
gain insight into the formation of the highly-sheared strong-field
polarity inversion line involved in the onset of the flare and coronal
mass eruption. Based on these simulations, they argue that the region,
with a surface appearance of two adjacent emerging bipolar regions,
was formed from a single subsurface structure that fragmented during
its rise \citep[see also][]{2013ApJ...764L...3C}. The interaction
between two opposite-polarity flux bundles in a glancing collision
created the strongly-sheared high-$R$ PIL conditions \citep[see
also][]{2013SoPh..288..507W} that they suggest enabled the formation
of a flux rope that appeared to play a key role in the observed flare
and CME \citep[see also][]{2015ApJ...803...73I}.

The existence of such a flux rope is consistent with work by
\cite{2014ApJ...783..102M} who use the coronal loops observed by {\em AIA}
in combination with the surface magnetic field to compute a NLFF
field. They note: ``Immediately prior to the eruption, the model field
contains a compact sigmoid bundle of twisted flux that is not present
in the post-eruption models''. The study by
\cite{2013SoPh..287..415P} shows that the high-$R$ PIL, and in
particular structures around the end points of the inferred flux rope,
are locations where the horizontal field component increased
\citep[also noted by][]{2012ApJ...748...77S}, and the Lorentz force
shows the most pronounced change when comparing pre- and post-flare
vector magnetograms \citep[also seen in flares in ARs\,11166, 11283,
and 11429 described in subsequent sections,
see][]{2012ApJ...759...50P}

The sunspot linked to the X2.2 flare was rotating, as was
another adjacent to an M2.2 flare on the day before:
\cite{2015SoPh..290.2199L} note that the M2.2 flare occurs when the
spot rotation rate peaks.

\cite{2015ApJ...811...16K} estimate Poynting fluxes in a 6-day
sequence of photospheric vector-magnetic maps observed by {\em SDO}/{\em HMI} to
conclude that the estimated free energy of $\approx 2\times
10^{32}$\,erg is compatible within better than a factor of two with
the energy needed for the X-class flare on 2011/02/15 and with other
estimates of the free energy (see references in their paper).

\begin{figure*}
\begin{center}
\noindent\includegraphics[width=14.7cm]{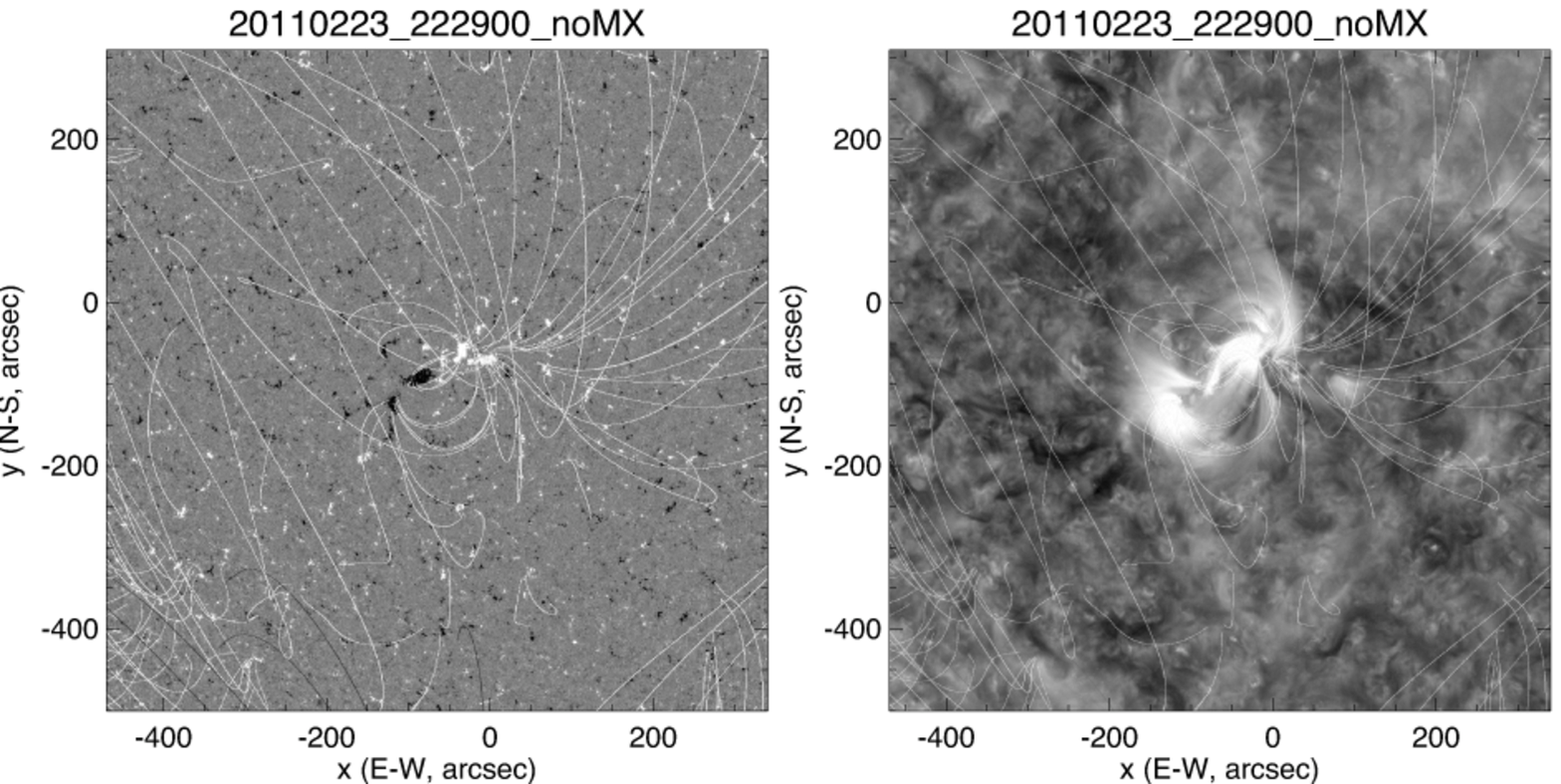} 
\noindent\includegraphics[width=14.7cm]{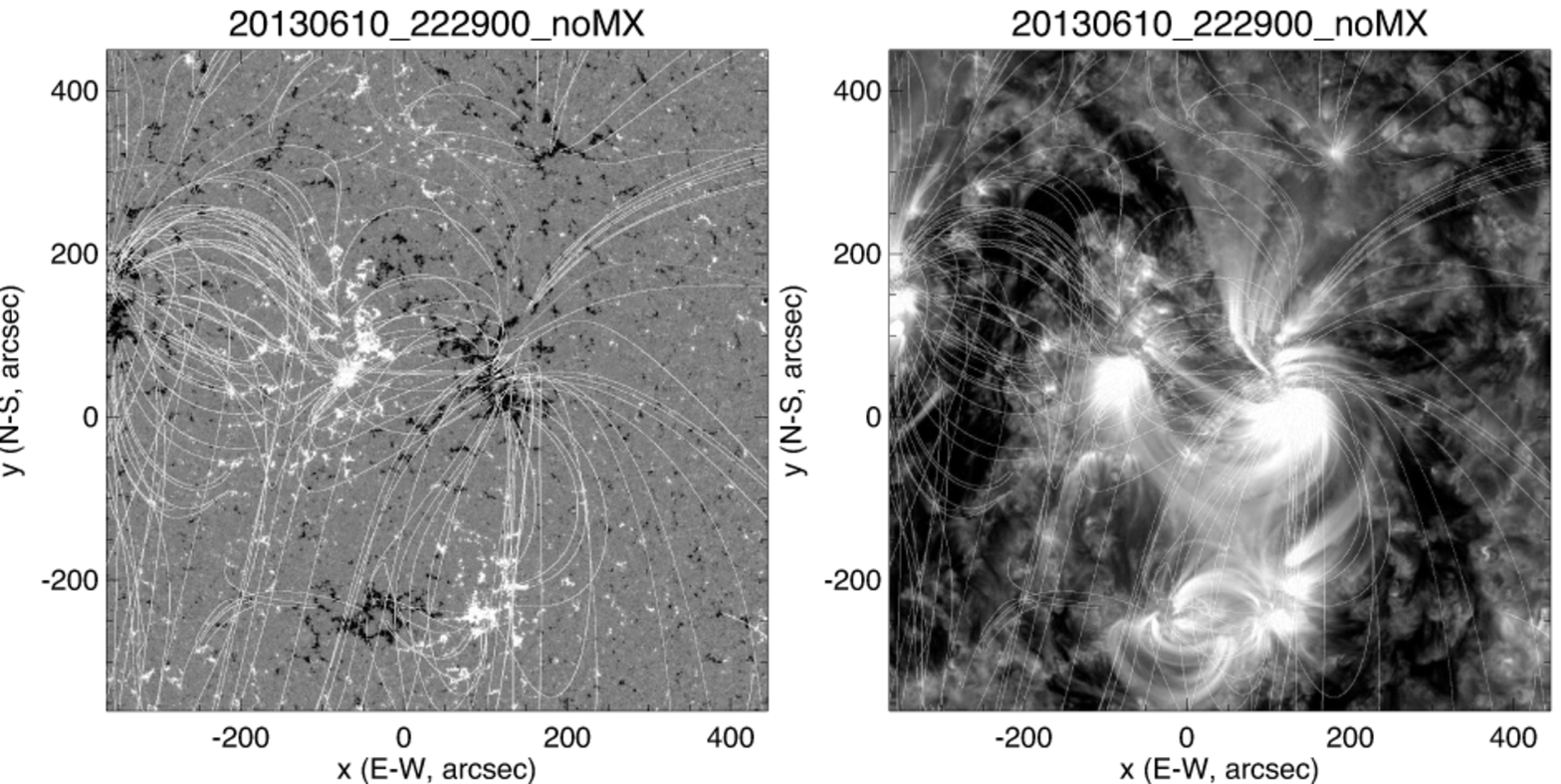} 
\end{center}
\caption{As Fig.~\ref{fig:3}: Unnumbered spotless regions, both without major flaring:
{\em (top)\,} region no.\ 4
{\em (bottom)\,}  region no.\ 29.
}\label{fig:6}
\end{figure*}
\subsection{Region no.\ 4} 
This unnumbered spotless region (located at $(x,y)=(-60, -96)$~arcsec
from disk center at the selected time) is shown in the top panels of
Fig.~\ref{fig:6}. It has a counterclockwise spiral from its leading
polarity flux deviating strongly from the PFSS model field. There are
no signs of flux emergence in the four days leading up to the time
selected for this study (picking up the region at 60 degrees
East). The {\em SDO}/{\em AIA} summary movies$^[2]$ show no substantial eruptive
activity on the date selected.

\subsection{AR\,11166} 
The active region-scale loops in the {\em AIA} 171\,\AA\ image are not
unambiguously nonpotential, so this region was not selected based on
its nonpotentiality, but was added for review because of the X-class
flare within it. Some loops within the deep active-region interior do
appear to reflect nonpotentiality.

\cite{2012ApJ...761...86V} describe the magnetic evolution of
AR\,11166 (and of 11158) in the days leading up to the X-class flare,
describing the shearing caused by the relative motions of the opposite
polarity flux concentrations.  See \cite{2014ApJ...792...40V} for a
description of the X1.5 flare and NLFFF field modeling.

The magnetogram sequence suggests that the high-$R$ PIL in the
interior of the region formed largely as a result of the displacement
of emerging flux in the SE periphery of existing flux, with the
leading polarity moving towards the west and eventually northward,
compressing and shearing opposite polarities.

%\subsection{AR\,11176} 
%The region, shown in the bottom panels of Fig.~\ref{fig:1}, has clear
%signs of nonpotentiality in southward loops from the interior of the
%leading polarity southward, as well as in a predominantly
%counterclockwise spiral from the leading spot. On 2011/03/28 there is
%only mild activity in the interior and towards the northeast, but on
%the preceding day, the region exhibited a large eruption from its
%northwestern extremity.

%\subsection{AR\,11236} 
%This region, shown in the bottom panels of Fig.~\ref{fig:4}, has
%limited nonpotentiality, essentially restricted to southward loops
%from the interior of the leading polarity.

%\subsection{East of AR\,11240} 
%The aged bipolar region leading AR\,11240 shows no spots or pores in
%the {\em SDO}/{\em HMI} images. AR\,11240 begins to emerge at the start of
%2011/06/20, and continue to add flux for through the end of the 23rd
%(displaying half a dozen pores). It has a counterclockwise swirl from
%the leading polarity.

\subsection{AR\,11283} 
There is apparent nonpotentiality in the interior and likely on the
northwest perimeter of the region, but not obviously so in the
active-region scale loops. 

Flaring is connected to compact flux emergence north of the leading
sunspot. The X1.8 flare on 2011/09/07 is notable because of a
substantial amount of chromospheric material being ejected. The field
evolution around the time of the two X-class flares was described by
\cite{2014ApJ...784..165R,2015ApJ...812..120R}. These studies note
that both major flares were associated with clear sunspot rotation. The SHIL
involved in the flaring started to form with flux emergence from about
2011/09/03 09\,UT onward, i.e. for some 3.5\,days prior to the first
X-class flare. The flux emergence, on the north-west side of the
leading sunspot, pushed the trailing, positive (rotating) polarity
into pre-existing spots, while remaining connected to the leading
polarity moving ahead of the region while connected by a filament and
--~in the NLFFF model~-- a flux-rope
structure. \cite{2015ApJ...812..120R} note that whereas the polarities
of the intruding bipole continue to separate, ``no apparent flux
emergence is observed during the period between the X2.1 and X1.8
flares,'' and that a filament configuration persisted after the X2.1 event.

\subsection{AR\,11289/-93} 
The {\em SDO}/{\em AIA} observations suggest a large eruption from these regions
in the final hours of 2011/09/13, but with no major flaring, although
there are apparent post-eruption loops for the first $\approx 6$\,hrs
of 2011/09/14. The loops arching equatorward are most inconsistent
with the PFSS field model.

%\subsection{AR\,11330} 
%The region is most notably nonpotential in the clockwise swirl of
%loops over the leading polarity and in equatorward loops from the
%trailing polarity.
%
%There is a small bipolar region emerging northeast of AR\,11330
%starting 2011/10/25 after 17\,UT. That region appears to be eruption
%on 2011/10/28 after about 16\,UT, with some effects on the corona of
%AR\,11330. Otherwise of note is a large quiet-Sun filament
%configuration that wraps around the region from the northeast to the
%southwest, a large part of which erupts around midday on 2011/10/27.

\subsection{AR\,11362} 
The region, shown in the bottom panels of Fig.~\ref{fig:5}, has
apparent nonpotentiality in loops from its leading polarity that
connect to a decayed region to the south.

A mid-sized bipolar region emerges to the north, starting around
2011/12/01 21\,UT, ending its flux increase at the end of
2011/12/03. This causes coronal deformation in the
northern reaches of AR\,11362 around 2011/12/03 14\,UT, and more on
the 4th, with also what is likely an eruption from AR\,11362 around
16\,UT on that day.

%\subsection{AR\,11399} 
%The region is connected to a long quiet-Sun filament extending towards
%the northwest, where nonpotentiality is most notable. The field
%configuration associated with that activates at the end of 2012/01/17,
%continuing to evolve until the middle of the 18th, then erupts (during
%an {\em SDO} off-point maneuver) exhibiting large post-eruption loops that
%continue to glow until past the end of the selected date.

\begin{figure*}
\begin{center}
\noindent\includegraphics[width=14.7cm]{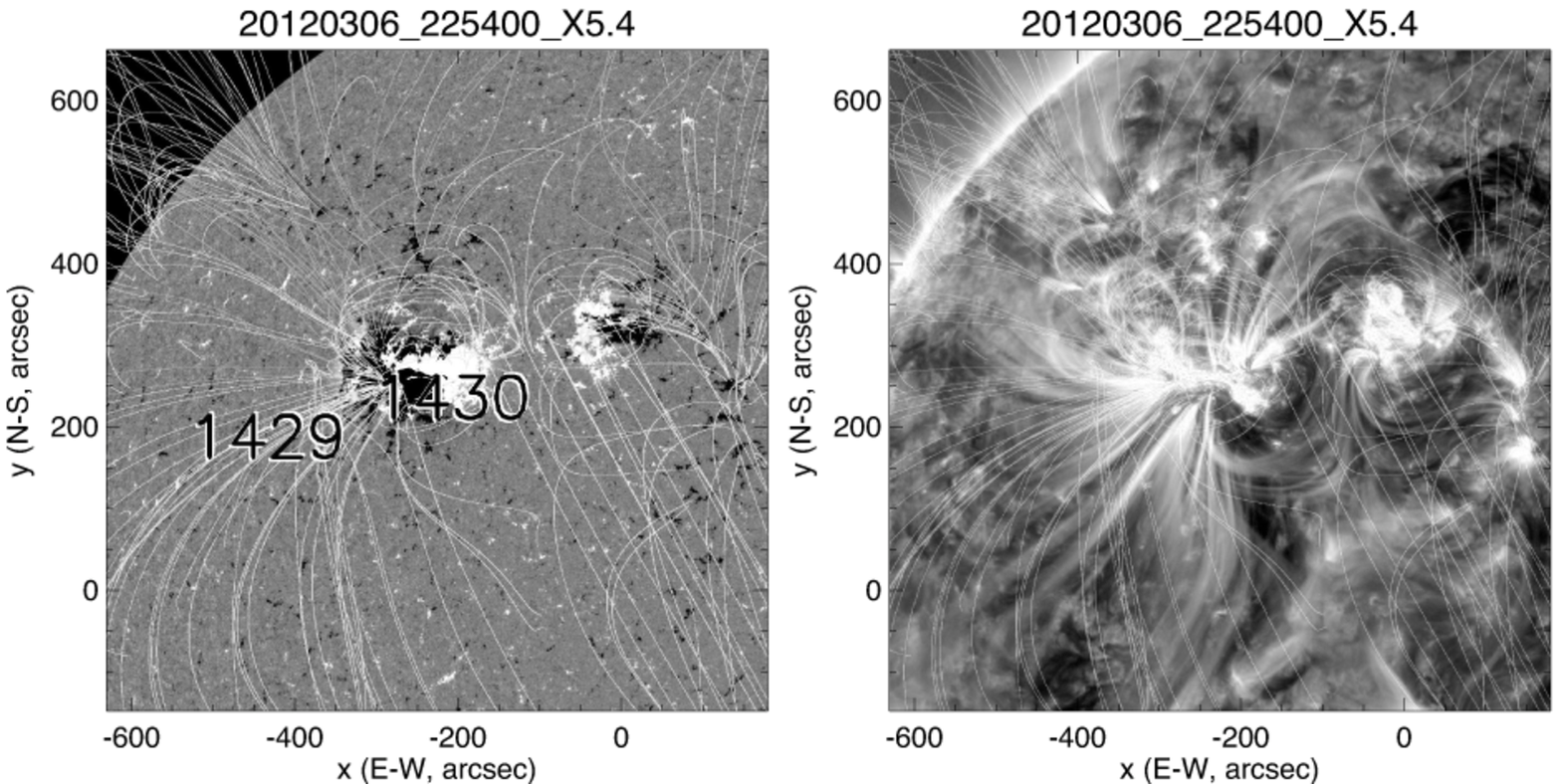} 
\noindent\includegraphics[width=14.7cm]{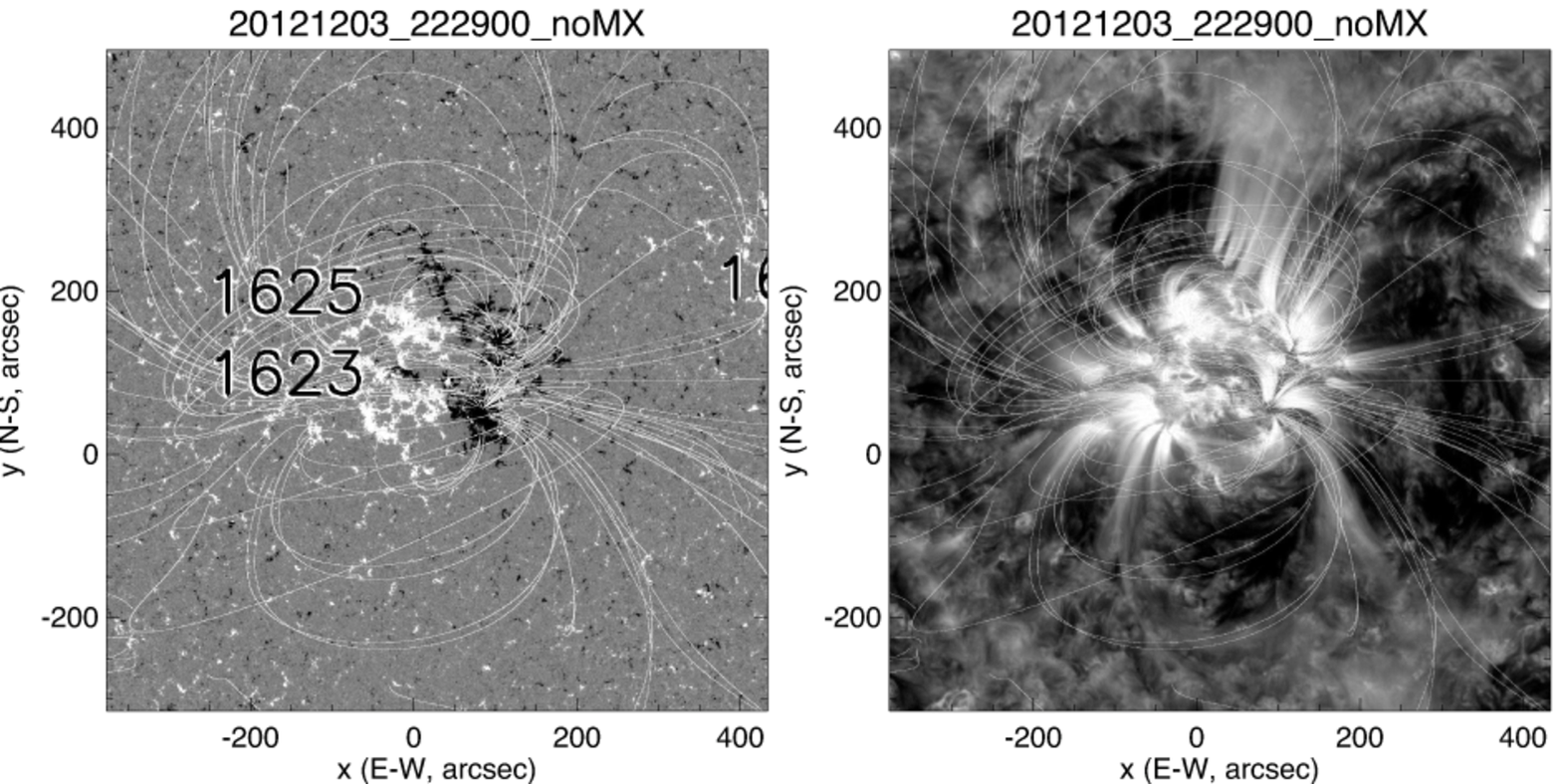} 
\end{center}
\caption{As Fig.~\ref{fig:3}: Example regions in category 'b':
{\em (top)\,} AR\,11429/-30
{\em (bottom)\,}  AR\,11623/-25.
}\label{fig:2}
\end{figure*}
\subsection{AR\,11429/-30} 
The region (Fig.~\ref{fig:2} top) exhibits
nonpotentiality in connections to a decayed region to the south as
well as  towards the north, and likely in loops emanating from its
interior. Note also the counterclockwise swirl from the trailing
polarity of the region just ahead of it, AR\,11430.

Leading AR\,11429 is a newly emerging flux region, starting around
2012/03/04 05\,UT, which becomes AR\,11430.  AR\,11429 exhibits
frequent activity, both within the region, and in the evolving
connections to AR\,11430 on 2012/03/06. There appears to be much less activity on
2012/03/08 (but note that {\em AIA} coverage on 2012/03/08 misses multiple
hours), but there is a major eruption on 2012/03/08 after 03\,UT. 

Internally, AR\,11429 has a long SHIL, being maintained for the
selected date range by sustained shearing and converging displacement
of the fluxes around it.

\cite{2015ApJ...809...34C} describe the evolution of AR\,11429 from
2012/03/05 to 2012/03/08, including the evolution of a rope structure
above the SHIL, the sunspot rotation, coronal activity,
and two fast CMEs (exceeding 2000\,km/s) in essentially opposite
directions that occurred early on 2012/03/07 in association with the
X5.4 and X1.3 flares. Their NLFFF extrapolation for late on the
2012/03/06 contains a weakly-twisted flux rope configuration along the
length of the primary PIL that they argue is composed of two segments,
each one of which involved in a separate flare and related CME. 

%\subsection{AR\,11442} 
%This region has apparent nonpotentiality in loops from this region
%connecting to a trailing compact bipole.
%
%AR\,11442 started emerging in the early hours of 2012/03/25, adjacent
%to a pre-existing, mid-sized, aged bipolar region to its
%northwest. Flux emergence intensified early on 2012/03/26, along with
%another bipolar region due east of it. The pre-existing region flared
%on 2012/03/28 around 22\,UT.

%\subsection{AR\,11470/-1/-2} 
%A triplet of regions, with only AR\,11471 containing a large sunspot
%in its leading polarity. No substantial flux emergence occurred during
%the 4 days prior to the time selected for this study.
%nonpotentiality is likely in equatorward loops from both regions.

%\subsection{AR\,11490} 
%Some nonpotentiality is suggested in loops connecting from its
%leading polarity to a leading decayed region. Some flux is still
%emerging within the region as a new substantial bipole begins to
%emerge around 2012/05/27 08\,UT at its southeastern edge.

%\subsection{AR\,11497} 
%There is possible nonpotentiality in the clockwise swirl of loops
%emanating from the trailing polarity. The polarities continue to
%separate as emergence phase is ending from 2012/06/01 onward.

\subsection{AR\,11519/-20/-1} 
This area is a complex of three substantial, closely-packed active
regions, with flux emergence throughout, which hampers the assessment
of is nonpotentiality.  Loops on the northern, trailing side of
AR\,11520 are inconsistent with the PFSS field lines, as are low-lying
chromospheric (absorption) features in the 171\,\AA\ image. The
leading AR\,11521 does not start to emerge until $\approx 15$\,UT on
2012/07/08 continuing until about 2012/07/09 18\,UT. The main SHIL in
AR11519 is sustained by converging motions of the main flux clusters.

%\subsection{AR\,11532/-36} 
%With possible nonpotentiality in southward loops from the central
%area.  Flux emerges into the trailing polarity of AR\,11536 with rapid
%separation between the polarities. Flux emergence weakens about about
%2012/07/29 14\,UT.

%\subsection{AR\,11542}
%There is a counterclockwise swirl in the loops from the leading
%polarity that appears nonpotential. A large eruption occurs towards
%the south around 16:30\,UT, with a sustained afterglow in the coronal
%loops until at least around 22\,UT. [Possibly the field is still
%evolving]

\subsection{AR\,11555/-60/-61} 
This is a group of bipolar regions in which AR\,11561 begins to emerge
on 2012/08/29 around 06\,UT until the moment selected for this study,
with polarity mixing in its interior.  There is a cluster with
multiple internal loop sets that is suggestive of nonpotentiality.

A very large and extended quiet-Sun filament eruption occurs towards
the southeast of the region, of which the leading end connects to
loops from AR\,11561. That region exhibits a possibly induced eruption
on its leading edge around 20:30\,UT on that day, followed by another
at 22\,UT, then again on 2012/09/01 around 01\,UT, 05:30\,UT, 09\,UT,
and 12:30\,UT, with another from the region to the north around
18\,UT.

%\subsection{AR\,11569/-71/-74} 
%ARs\,11571 and 11569 are perfectly aligned in the east-west
%direction. AR\,11574 begins to emerge towards the southwest around
%05\,UT on 2012/09/16.  There is apparent nonpotentiality of loops
%emanating from interior polarities of the complex.

%\subsection{AR\,11599} 
%AR\,11599 at the selected time is essentially a unipolar region with a
%large spot of the leading polarity in the southern hemisphere, as it
%was for at least the preceding four days.  There is a pronounced
%counterclockwise swirl of loops from the (formerly leading) spot (with
%opposite polarity apparently cancelled or strongly dispersed).

%\subsection{AR\,11621} 
%There is some signature of nonpotentiality in the clockwise swirl of
%loops from the leading polarity in this region with well-separated
%polarities: AR\,11621 is a mature region, with the leading spot
%essentially containing the leading flux, and with the trailing
%polarity distant and dispersed. There is the common moat activity
%around the leading spot, but involving only relatively little flux.

\subsection{AR\,11623/-25} 
ARs\,11625 and 11623 are a pair of comparable sizes, perfectly aligned
next to each other, 11625 to the north of 11623. On 2012/11/29,
AR\,11625 is still emerging, but by the end of that day, polarities
are separating.  The region, shown in the bottom panels of
Fig.~\ref{fig:2}, is most clearly inconsistent with the PFSS field
model in loops emanating from the leading polarity of the northern
region (AR\,11625).

%\subsection{AR\,11633/-4} 
%Two adjacent regions, both fully developed, with AR\,11633 leading
%11634 by half a polarity separation. There are multiple moderate
%eruptions from the leading spots of AR\,11633, arcing along loops on
%the northern side, with the main ones around 2012/12/19 02\,UT,
%10\,UT, 18\,UT, and 24\,UT.  A counterclockwise swirl in loops from
%the northern leading spot region is most suggestive of
%nonpotentiality.

%\subsection{AR\,11745} 
%There is possible nonpotentiality from flux in the southern trailing
%polarity.  \cite{2013ApJ...778L..29L} describe a series of eruptions
%of ``homologous flux ropes'' during the period of May 20-22, 2013.

\subsection{Region no.\  29}
This region, located at $(x,y)=(42,45)$~arcsec from disk center on
2013/06/10 (shown in the bottom panels of Fig.~\ref{fig:6}), has a
strong clockwise swirl in the loops from its leading
polarity that differ markedly from the PFSS model field. It is a
decaying bipole, with both polarities broken up in supergranular flows,
and well separated.

\subsection{AR\,11864/-5} 
This region has possibly nonpotential loops in its interior.  There
is an eruption from the trailing part of the flux complex around
00:30\,UT, and a coronal evolution that could be another eruption
around 20\,UT with loops continuing to reshape until beyond the end of
the day on 2013/10/13, then another eruption on 2013/10/14 around
13\,UT, and one from the leading flux cluster at $\approx$16:30\,UT
with loop deformations continuing until past the end of the 14th.

\subsection{AR\,11890} 
The region (Fig.~\ref{fig:3}, top) has no 
obvious signs of nonpotentiality in its overall appearance, 
although the leading spot shows signs of a counterclockwise swirl.

The X flares occur over compact flux emergence with strong
shear flows at the trailing end of the region, with both flares
displaying \correction{pair of nested post-eruption loop arcades as expected for the 
multipolar area involved in the flares.} Loops around
that area do not obviously disagree with the PFSS model.

\subsection{AR\,11895/-7} 
There are possible signs of nonpotentiality in loops from the leading
polarity of AR\,11895.

Shearing of a high-$R$ region appears to exist when AR\,11895 is near
the limb, but this fades early on the 12th; by the 14th the region has
largely separated polarities, with a large leading spot well ahead of
the remainder of the leading polarity flux that is already largely
fragmented, as is the trailing polarity.

The region shows no substantial activity until an eruption on its
northern side on 2013/11/16 around 08\,UT with subsequent loop
evolution lasting until the end of the day. On the 17th, continuing
subdued on the 18th, there is
frequent small-scale activity with jet-like eruptions from the
southern side of the trailing polarity, likely associated with flux
emergence into the area. 

\begin{figure*}
\begin{center}
\noindent\includegraphics[width=14.7cm]{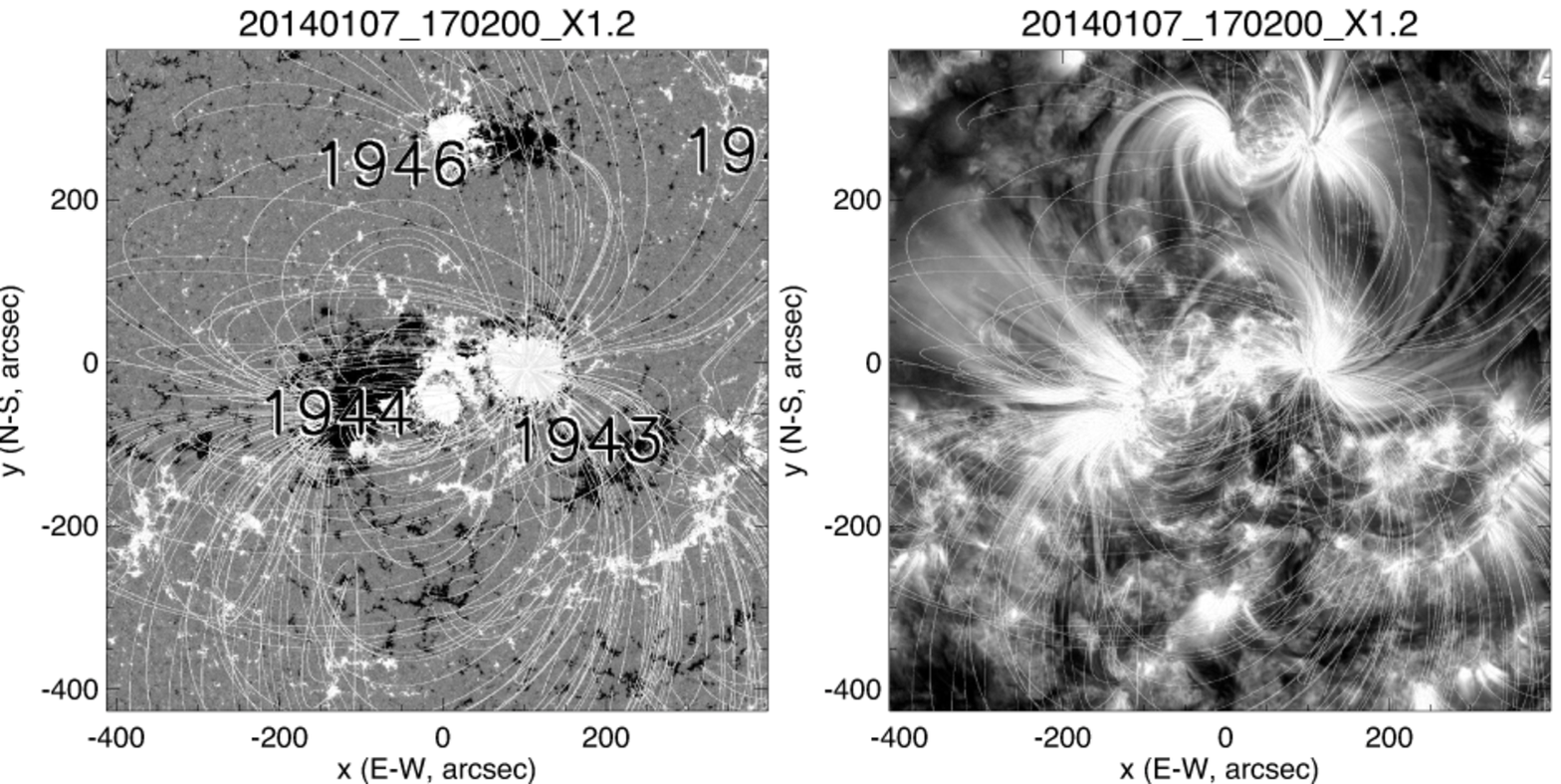} 
\noindent\includegraphics[width=14.7cm]{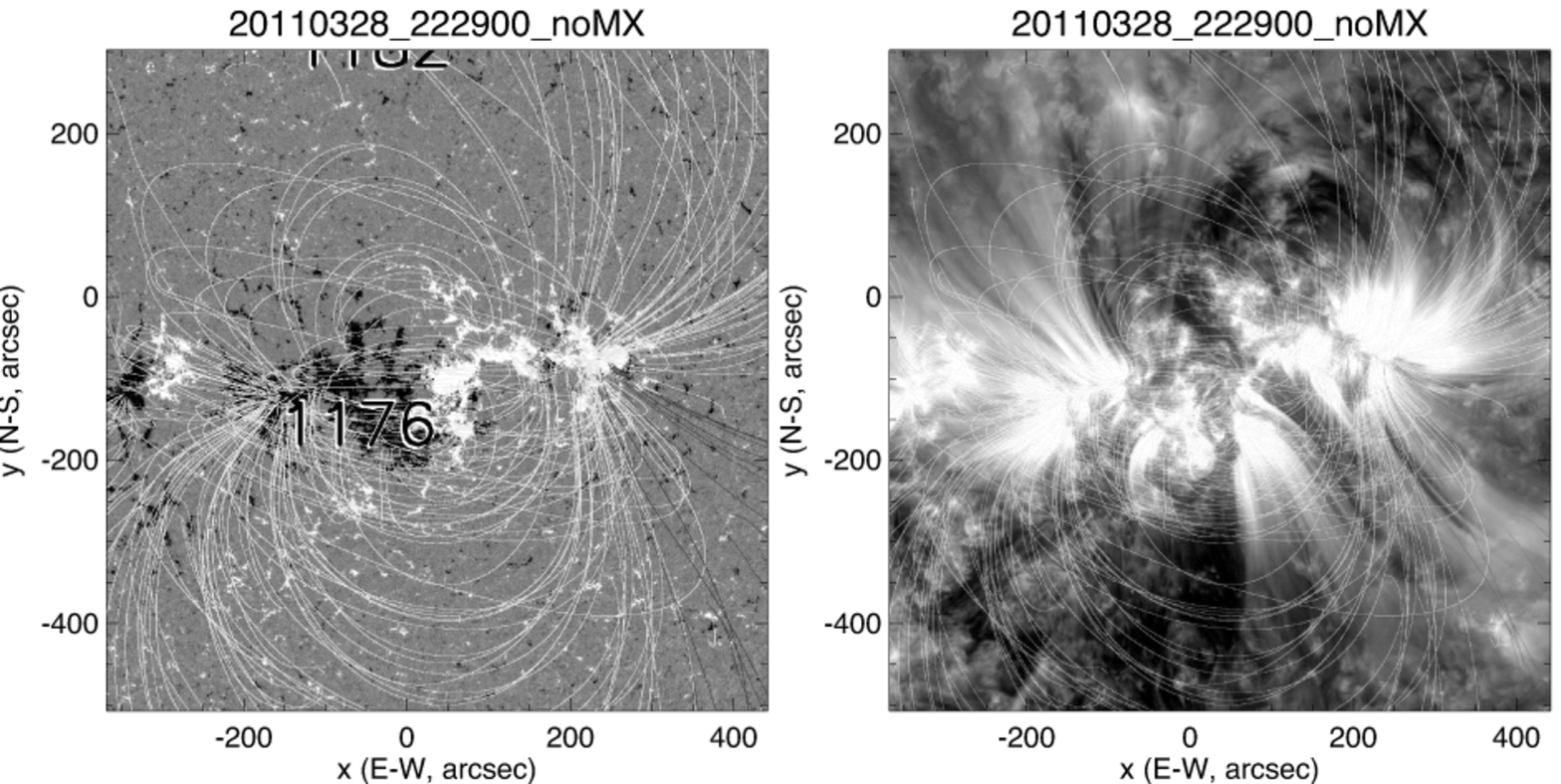} 
\end{center}
\caption{As Fig.~\ref{fig:3}: Example regions in category 'a', with and without X flares:
{\em (top)\,} AR\,11944
{\em (bottom)\,}  AR\,11176.
}\label{fig:1}
\end{figure*}
\subsection{AR\,11944 and AR\,11946} 
The flaring region, AR\,11944 (Fig.~\ref{fig:1}, top)
is not obviously nonpotential, although the loops
from the leading spot shows a very pronounced counterclockwise
swirl. Loops in the core region of AR\,11944 also exhibit
nonpotentiality.  Moreover, the neighboring small AR\,11946
is nonpotential, judging from the counterclockwise spiral of its
trailing polarity.

There are multiple high-gradient regions contributing to the region's
$R$ value, reflecting compact flux emergence, convergence, and
cancellation in the region's interior along the overall PIL. Northward
of the region, a new bipolar region begins to emerge round 2014/01/03
22\,UT, continuing throughout the next few days; emergence ends around
2014/01/08 when the polarities are separated around 03\,UT. 
This emerging region, AR\, 11946, is well connected to AR\,11944, in particular
visibly to the leading sunspot from which loops exhibit a
counterclockwise swirl. The X flare originates from under a loop
system to the south of that. This is an unusual eruption of field
connecting a large leading spot with dispersed trailing field of old
AR on the leading side. {\em AIA}\, 304\,\AA\ observations suggest filament
involvement. Note that no high-$R$ region appears to be connected to
the X flare.

%\subsection{AR\,11950} 
%There is a curved structure or preferential swirl from the trailing
%polarity of this region suggestive of nonpotentiality.
%
%The region is still showing signatures of flux emergence, with
%polarities increasing their separation, and some shearing of the field
%until at least early on the 12th.
%
%A large QS filament trailing the region lifts off starting around
%05\,UT on the 13th, with eruption afterglow at least until 16\,UT,
%with an arcade connecting into the trailing polarity of AR\,11950
%where the signature of nonpotentiality is strongest. 

%\subsection{AR\,11970} 
%There is nonpotential swirl in the internal loops in this small
%region in connections between the polarities. It has mostly separated
%polarities, but with some flux emergence at the PIL and in the
%trailing polarity. There appears to be a slow eruption around 21\,UT.

%\subsection{AR\,12017} 
%The 171\,\AA\ loops are incompatible with the PFSS model only for
%low-lying interior loops.
%
%Flux forming the high-gradient strong-field  PIL begins to emerge at
%the leading spot around 2014/03/27 16\,UT continuing to increase $R$. 
%
%There is a large eruption associated with strong flaring on the
%leading side of the region on 2014/03/28 $\approx 19$\,UT, again just
%before midnight, and once more on the 29th $\approx 18$\,UT.

\subsection{AR\,12071/-3} 
AR\,12071 is largely complete showing little flux emergence, but
AR\,12073 has strong flux emergence for several days intruding into an
existing bipole. The leading spot of the latter persists, and spot and
flux of the new emergence move towards it and appear to be deflected
southward just prior to merging with it.  There is a nonpotential
swirl over the leading polarity of AR\,12071.

%\subsection{AR\,12090} 
%Mature moderate-sized region with well-separated polarities. The
%leading polarity is essentially concentrated in a spot (with some moat
%activity), and the trailing polarity is dispersing in the
%supergranular flows.  There is a clockwise swirl in loops from the
%leading polarity.

\subsection{AR\,12158} 
This is a strong, compact region with most likely moderate
nonpotential loops particularly from the southern/trailing polarity
(suggesting a clockwise swirl), as well as for loops from the leading
edge of the southern/trailing polarity. The X-class flare involves a
long, curved filament riding on the PIL, clearly indicating sheared
field, involving a few relatively small SHILs in three patches along
its length. There is considerable activity and emergence in its moat
region, particularly towards the south, pushing into the trailing
polarity.

\begin{figure*}
\begin{center}
\noindent\includegraphics[width=14.7cm]{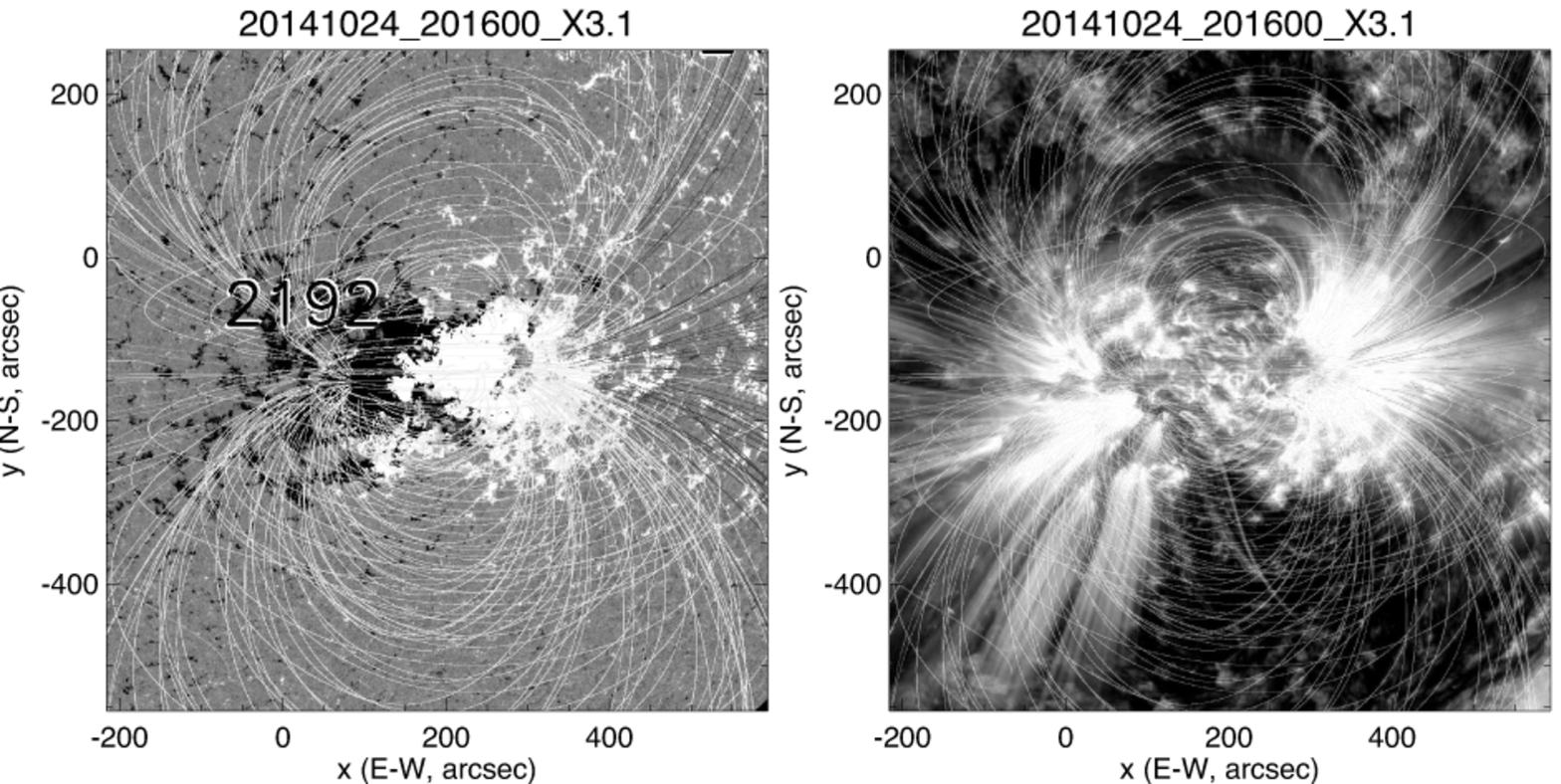} 
\noindent\includegraphics[width=14.7cm]{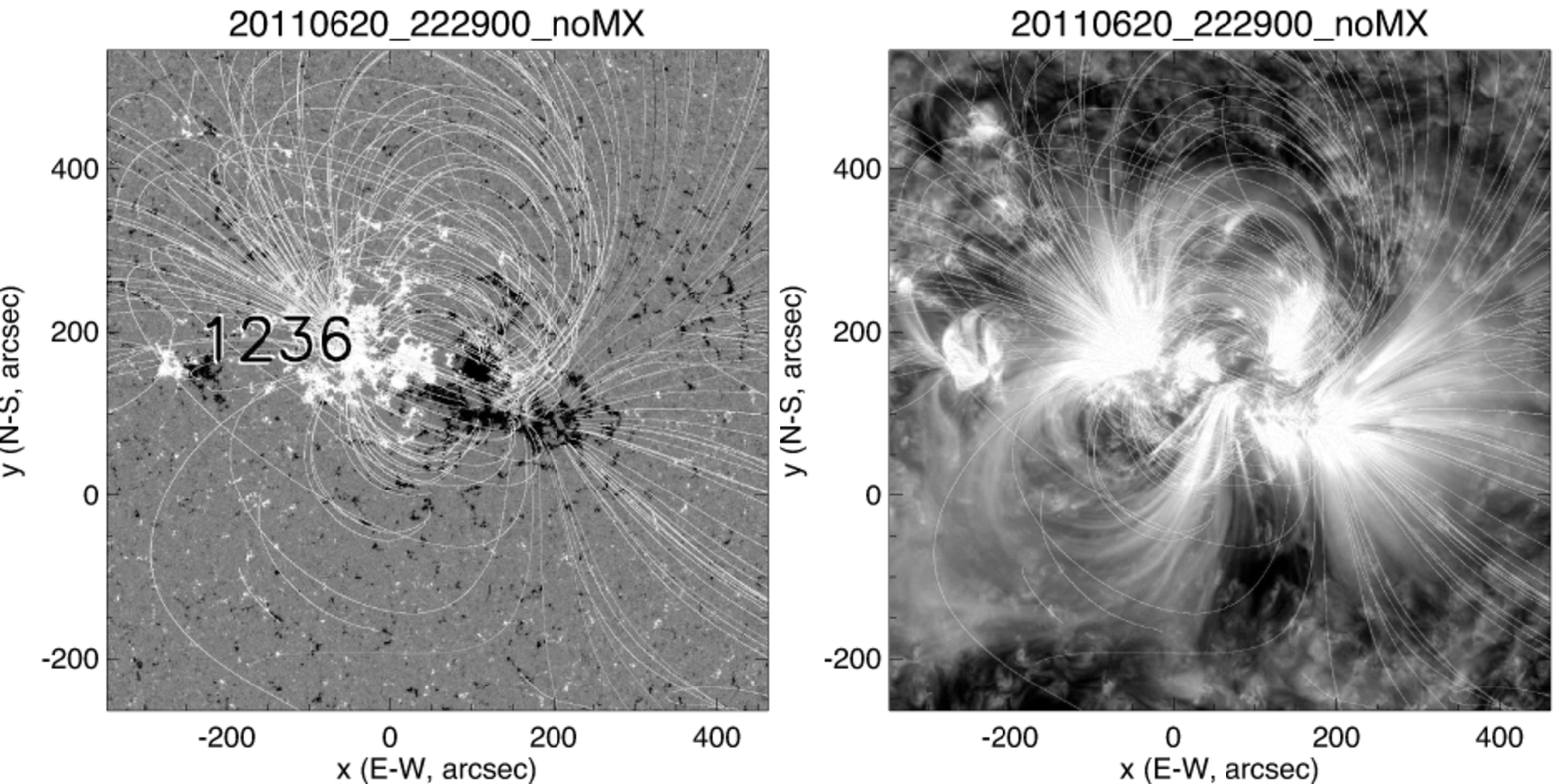} 
\end{center}
\caption{As Fig.~\ref{fig:3}: Example regions in category 'd', with and without X flares:
{\em (top)\,} AR\,12192
{\em (bottom)\,}  AR\,11236.
}\label{fig:4}
\end{figure*}
\subsection{AR\,12192} 
AR\,12192 is a very well studied region, attracting considerable
attention because it exhibited a series of 6 X-class flares none of
which had a corresponding CME. There were another 29 M-class flares
of which only one, originating from the periphery of the region (where
5 other M flares occurred), was associated with a CME
\citep{2015ApJ...808L..24C}. See \cite{2015ApJ...804L..28S} for
further discussion and interpretation of why this region exhibits so
many energetic confined flares.

At the time studied here, the region is mostly formed, but there is
much ongoing flux emergence, including shearing motions and densely
packed opposite polarities in at least four places scattered around
the main PIL. 

The region, shown in the top panels of Fig.~\ref{fig:4}, is clearly
nonpotential in some loops from its trailing polarity, and has
multiple chromospheric (absorbing) fibrils and filaments suggestive of
sheared field. But the appearance of the other active region-scale loops is
not unambiguously nonpotential, and even nonpotentiality in the core
appears limited \citep{2015ApJ...804L..28S}.

\section{Data analysis and review}\label{sec:analysis}
The observations summarized above show that high-energy X-class flares
and pronounced nonpotentiality as inferred from loops observed in the
{\em AIA} 171\,\AA\ or 193\,\AA\ channels are only moderately correlated: only six of
ten active regions with X-class flares exhibit clear signs of
nonpotentiality in region-scale loops, while clearly nonpotential
regions do not necessarily display significant flaring (or major
eruptive activity).

Examples of nonpotential regions without major flaring include
AR\,11176 (with $\log(R)=3.3$), AR\,11362 ($\log(R)=3.5$), ARs\,11569/-71/-74
($\log(R)=2.9$), AR\,11621 ($\log(R)=1.8-2.3$), AR\,11745
($\log(R)=2.3-2.6$), AR\,12071/-3 ($\log(R)=2.9$), and AR\,12090
($\log(R)=2.2-2.9$), as are the two spotless (and consequently
unnumbered) regions.

In some cases, nonpotentiality may be a temporary signature of a
preceding eruption from which the coronal field is still relaxing
(regions marked with an asterisk in Table~\ref{tab:summary}). The
post-eruption reconfiguration of the higher coronal configurations can
take many hours 
\citep[see, e.g.,][]{2013ApJ...773...93S}. In one such
case, AR\,11399, the loop system is reflecting the
afterglow of a large filament reconfiguration extending from the AR
into distant quiet Sun that started at the end of 2012/01/17 (with an
eruption that was not observed, occurring during one of the infrequent
{\em SDO} off-point maneuvers), and continuing loop evolution and afterglow
until past the end of the selected date. Similarly, AR\,11542 erupted
around 16:30\,UT on the selected date, with sustained afterglow of the
loops at least until around 22\,UT, so this region, too, may still be
relaxing from an eruption. Also AR\,11289/-93 exhibited a large
eruption in the final hours prior to the selected time, with afterglow
continuing for at least some 6\,h into the next
day. ARs\,11555/-60/-61 also are observed between eruptions, with one
preceding the selected time by only about half an hour. And
ARs\,11864/-5 appears to be still reconfiguring after an
eruption. Possibly even AR\,11970 is subject to the same: a slow
coronal change, reminiscent of an eruption, occurred around 21\,UT
with some continuining evolution for a few hours afterwards.

In cases like AR\,11519/-20/-21 the corona may be
very complex and dynamic and a moderate-resolution PFSS field model
could be inadequate to approximate the appropriate
potential-field model.
 
\begin{table}[t]
  \caption{Contingency table for $\log(R)$  measuring the unsigned SHIL flux
    associated with flaring in excess of M5 in the sample.}\label{tab:rcontingency}
\begin{center}
\begin{tabular}{l|cc}
\hline\hline 
No.\ of regions:& no M or X & M or X  \\
\hline
$\log(R) \le 4.0$:  &49 & 0 \\
$\log(R) > 4.0$:  & 11 & 18 \\
\hline
\end{tabular}
\end{center}
\end{table}
For other cases, in which the photospheric field is simply structured
and the corona not particularly dynamic, there is no obvious pattern
in an instantaneous magnetogram to infer the degree of
nonpotentiality.  Regions without large flares on a given day can appear
strongly nonpotential even when the polarities are moderately to well
separated; see, for example, AR\,11092 ($\log(R)=2.6$),
AR\,11236 ($\log(R)=2.4-2.9$), AR\,11362 on
2011/12/02-04 ($\log(R)=2.8-3.5$), AR\,11399 
($\log(R)=0$), AR\,11497 ($\log(R)=2.1$), and AR\,11542
($\log(R)=3.9$), or they may have a predominantly
bipolar configuration with some satellite regions (such as AR\,11330, 
AR\,11240 and preceding region, and the unnumbered
region on 2011/06/23).  Nonpotentiality for such regions may even be
evident in case one polarity has largely dispersed and cancelled,
leaving a predominantly single polarity region. The clearest example
of this is AR\,11599, but AR\,11621 and AR\,12090 are
to some extent comparable: their trailing polarity still exists, but
is distant from the large leading spots that contain essentially all of
the leading-polarity flux.  Other nonpotential regions have a
multipolar structure (such as AR\,11176, and
AR\,11895/-97). Some (e.g., AR\,11442,
ARs\,11470,-71,-72, ARs\,11555/-60/-61,
ARs\,11569/-71/-74, ARs\,11614/-16/-19)
show nonpotentiality in the connections between fairly-well separated
satellite regions, or even regions so dispersed they are essentially
enhanced network (AR\,11490).

\begin{figure}
\noindent\includegraphics[width=8.4cm]{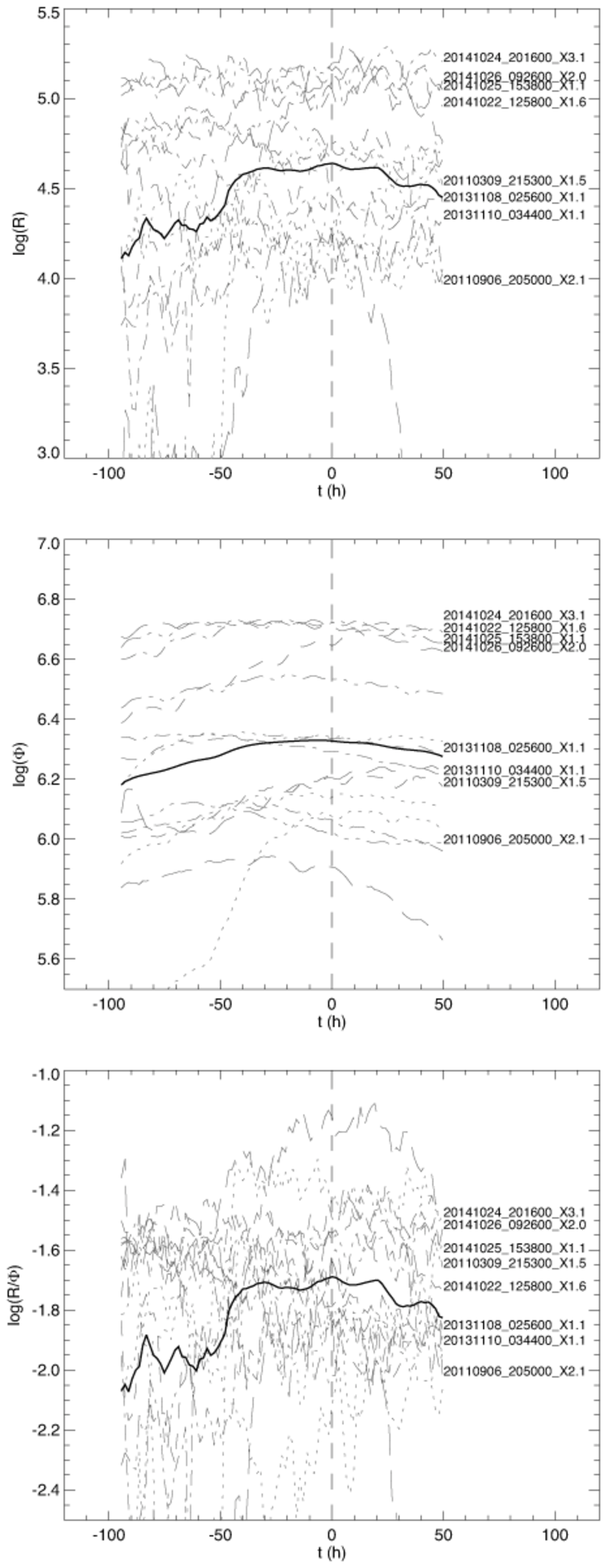} 
\caption{For regions with X-class flares, the superposed epoch
  diagrams show the values with time (on a 96-min.\ cadence) of
  $\log(R)$ (top), total absolute flux $log(\Phi)$ (in the same units
  of $2.2\times 10^{16}$\,Mx, see Schrijver [2007] for rationale;
  center), and the fraction of the total absolute flux involved in the
  strong-field high-gradient inversion line $\log(R/\Phi)$ (SHIL;
  bottom). The curves have been shifted to the flare time as
  reference, and are shown in grey using various line styles;
  labels identify regions in class c or d. The heavy curve is
  the average superposed-epoch trend for the set of 15 X-class flares,
  shown after a 5-point boxcar smoothing.}\label{fig:logrt}
\end{figure}
Dynamics of the field may be more telling: regions with magnetic field
patterns that have little recent flux emergence, well separated
polarities, simple bipolar configurations (i.e., categorized in
classes D-G defined in Sect.~\ref{sec:observations}) have $\log(R)$
values under 3.3, with one exception that comes in at 3.9. None of
these regions exhibited a flare at M5 or higher. In contrast, all
active regions with $>$M5 flares have values of $\log(R)$ exceeding
4.0 (see Table~\ref{tab:rcontingency}), and all have an increasing
value of $\log(R)$ or are within 0.1 of that peak at the time of the
major flaring; all but two have decreasing values of $\log(R)$ on
average over the subsequent two days (excepting only AR\,11520 and
-~weakly~- AR\,12192 on 2014/10/24); compare Fig.~\ref{fig:logrt} and
Table~\ref{tab:rtrend}.  Regions with X-class flares have
high-$R$-value interiors associated with flux emergence and
displacement, but many of these do not show obvious signs of
nonpotentiality in the "outer" 171\,\AA\ loops; one example is
AR\,11283 which shows possibly nonpotential
loops only in the area north of the leading spot (here the 6-h,
1-degree PFSS spatiotemporal resolution may be a limiting factor in
the analysis). Others (e.g., AR\,11429) have substantial
nonpotential signatures in their outer loops and a high $R$-value
interior (which, for this region, persists on subsequent days without
M- or X-class flaring); this is also the case for AR\,11944.

\begin{table}[t]
  \caption{Contingency table for the average 4-day trend in $\log(R)$ 
    ending in M- or X-class flaring or not. Only three cases were omitted from this table owing to erratic behavior in $R$, which was low in each of these. Trends are denoted as ``rising'' or ``falling'' if the change exceeded a factor of two in a linear fit, ``flat'' otherwise.}\label{tab:rtrend}
\begin{center}
\begin{tabular}{l|cccc}
\hline\hline 
Trend in $R(t)$: & falling & flat & rising & falling after X flare\\
\hline
no M or X flare & 36 & 19 & 2   & -  \\
M or X flare      & 1   & 6   & 10 & 13 of 15 \\
\hline
\end{tabular}
\end{center}
\end{table}
Active regions with $log(R) \ge 4$ without large flares all exhibit
activity within them or within their immediate vicinity: AR\,11330
which was close to, and possibly connected to, substantial eruptions
on the northern side; ARs\,11555/-60/-61 (which exhibit frequent
eruptions); ARs\,11864/-5 (with several large eruptions); and
AR\,11895/-7 (with eruptions and jets on 2013/11/17-18).

\begin{figure*}[t]
\noindent\includegraphics[width=\textwidth]{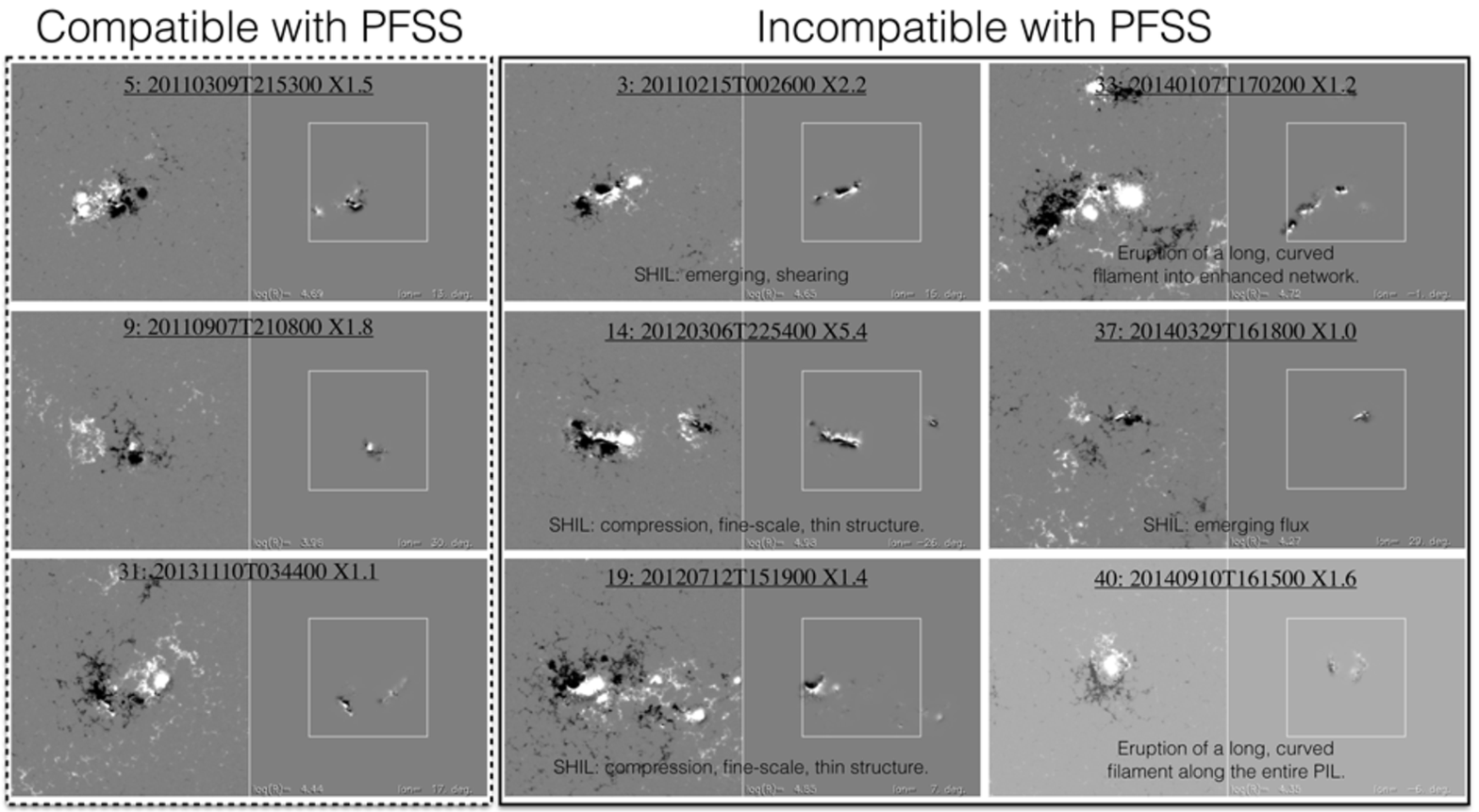} 
\caption{Comparison of {\em HMI} magnetograms and corresponding $R$ maps for regions with X-class flares; white lines encompass the 320-arcsec square areas within which $R$ is measured. Regions with largely potential coronae are shown on the left in the dashed box; others on the right in the solid box. AR\,12158 in which the X flare was not associated with a SHIL but with a long filament is shown in light gray. AR\,12192, with regionally pronounced  nonpotentiality, is excluded.}\label{fig:xpotential}
\end{figure*}
Three of the ten regions exhibiting M- or X-class flares do not
display obvious signs of nonpotentiality in their active region-scale
loops (classified as type ``c'' and/or ``d'' in
Table~\ref{tab:summary}): AR\,11166, AR\,11283, and AR\,11890 (shown
in the top panels of Fig.~\ref{fig:3}).  Nonpotentiality of AR\,12192
(Fig.~\ref{fig:4}, top) is particularly obvious in loops from its
trailing, southern area. The side-by-side comparison in
Fig.~\ref{fig:xpotential} suggests that a relatively long SHIL with
shearing motions along it is important for a pronounced nonpotential
appearance of active region-scale loops. Work by
\cite{2015ApJ...810...17D} showed that such shearing motions involving
the PIL induce non-neutralized currents in the surrounding corona,
thereby contributing to the nonpotential appearance. 

\correction{The three regions with near-potential coronae shown in
  Fig.~\ref{fig:xpotential} have SHILS off the primary PIL.  The
  primary SHIL for AR\,11890 at which the X1.1 flares occur is quite
  short and is located at the outer periphery of the active region.
  These properties may limit the influence of its current system on
  the region-scale loops. For the other two regions, the field
  geometry suggests that oppositely-directed mostly-parallel currents
  should be induced, thus weakening any effects on active-region scale
  loops. Specifically:}

\correction{The flare in AR\,11166 is associated with various clusters of mixed
  parasitic polarity that are caught in flows that push the various
  concentrations towards and along each other (see top-left panel in
  Fig.\,\ref{fig:xpotential}). The most pronounced SHIL just prior to
  the X1.5 flare is formed by a thin ridge of positive polarity
  sandwiched between negative polarities. In such a configuration, the
  velocity shear on the opposite sides of the positive-polarity ridge
  would be expected to induce oppositely-directed currents that would
  have largely canceling effects on the surrounding distant field.  A
  comparable configuration and dynamics were described by
  \cite{2014ApJ...781L..23W}, who showed the oppositely-directed
  magnetic twist in a non-linear force-free field model for AR\,11515
  (their Fig.~2c, with a map of vertical currents in their
  Fig.\,7i). In the X1.5 flare studied here, a second branch of the
  flaring area comprised a less pronounced, smaller, comparable
  configuration with negative polarity sandwiched between positive
  polarities (see \cite{2014ApJ...792...40V} for a discussion and
  NLFFF modeling of this event).}

\correction{The X1.8 flare in AR\,11283 is associated with a positive-polarity
  spot and surroundings being pushed into two negative-polarity
  patches on opposite sides, both containing spots with partly absent
  penumbrae \citep[see, e.g., Fig.\,2 in][]{2015ApJ...812..120R}. As
  for the X1.5 flare in AR\,11166, one can argue that
  oppositely-directed shear-induced current systems form between the
  three flux clusters.}

\correction{Returning to the non-potential regions,} Fig.~\ref{fig:xpotential}
suggests that the length and location of the SHIL relative to the bulk
of the region are important to nonpotentiality of the corona, but the
magnitude of $R$ will also play a role. The values of $R$ for the
selected regions show too little range to confirm this
observationally. Moreover, as shown in Fig.~\ref{fig:logrt} (bottom
panel), the fraction of the flux involved in the SHILs for the largely
potential X-flaring regions shows a spread comparable to the sample of
all X-flaring regions, i.e., somewhere between 1\%\ and 6\%\ of the
total absolute flux \citep[note that the $R$ metric involves only a
fraction of the flux that is involved in flux emergence as it was
quantified by][]{schrijver+etal2005a}.

I note that the X-class flare from AR\,11944, although in a region
with sunspots, does not involve a region with a SHIL.  This is the
only flare in this study not directly overlying at least in part a
high-$R$ area. This is discussed somewhat more in the next section.

Over the period of this study, {\em SDO}/{\em AIA} observed all active regions on
the disk, 1139 in total. Within the longitude selection window for
this study of about 80$^\circ$, some of these will not yet have
emerged, while others will have their spots (and with it their NOAA
number) disappear prior to entering the selection window. Assuming a
characteristic active-region life time of order ten days, regions
visible in the selection window will have emerged up to about 6 days
before entering the selection window, or $\approx$80$^\circ$ behind
the East limb. Regions can be given numbers to close to the West limb, so the total 
longitude range for numbered regions is $\approx 250^\circ$.
The fraction of NOAA regions selectable here is thus $\approx$210/250,
or somewhat over 80\%\ of the total.

Of these $\approx$950 regions, only 37 were selected as substantially
incompatible with the PFSS field, 5 of which may have their coronal
configuration still distorted while relaxing from eruptions that
occurred hours before the times of study, with another two regions
being unnumbered bipolar regions.  Consequently, only some $P_{\rm
  N}\approx$3-4\%\ of the sampled active regions appears nonpotential,
at least within the roughly 5-day window during which such regions
meet the selection criteria used here. Although the sampling is in
principle unbiased as I reviewed all {\em AIA} data within about
40$^\circ$ of disk center, the subjective selection criterion used
means this fraction should be interpreted with care. The sampling by
\cite{schrijver+etal2005a}, in contrast, in which roughly equal
numbers of the {\em TRACE} active-region pointings showed
near-potential and nonpotential coronae, was similarly subjective but
heavily biased by the pointing selections, often guided by community
members interested in AR activity. The value of $P_{\rm N}$ lies
well below the upper limit estimated for the fraction of ``less than
30\%'' of the lifetime of active regions during which flaring is
statistically observed \citep{schrijver+etal2005a}.

In some of the nonpotential regions, no significant flux emerges
during at least the 4 days preceding the time of the analysis
(although there are the usual moving magnetic features in the moats of
remaining sunspots, if any). These include all the regions with
essentially zero $R$ values: five numbered regions (ARs\,11084, 11092,
the region leading 11240, 11399, the essentially isolated spot region
11599), and the two unnumbered, spotless target regions Nos.\,4 and
29.  All have filaments in or next to them -~signatures of current
systems~- but then so do most sample regions;
\cite{schrijver+etal2005a} found no differentiating filament
characteristics between potential and nonpotential regions.

\cite{schrijver+etal2005a} noted that they could not attribute the
nonpotentiality to recent or ongoing flux emergence in 3 of 21
regions. Here, I find that 5 in 41 have had no flux emergence in at
least the four days preceding the selected time, which is
statistically un-differentiable from the number of nonpotential
emergence-free regions in \cite{schrijver+etal2005a}. Apparently,
nonpotentiality occurs for regions without (recent) flux emergence in
some 10\%-15\%\ of cases; the two nonpotential spotless regions
suggest nonpotentiality can exist even if the spotless region is
dispersing into the network.

\section{Discussion and conclusions}\label{sec:discussion}
This study uses 4.5 years of {\em SDO} observations to provide insight
into the origin and role of electrical currents in
active-region coronae.  As part of this work, I look into how to
integrate the conclusions from (a) \cite{schrijver+etal2005a} who
showed that regions with a distinctly nonpotential appearance of
their coronae were -~as an ensemble~- twice as likely to flare, with
average flare energies three times higher than for regions with
largely potential coronal fields, and (b) from \cite{schrijver2007}
that large flares require substantial amounts of flux in strong-field,
high-gradient polarity inversion lines (SHILs; expressed in a measure
denoted $R$). With this study, the sample of nonpotential regions
studied by \cite{schrijver+etal2005a} is increased from 36 to 77,
while the selection of the regions is free from the researcher-interest
bias that guided the {\em TRACE} target selections.

\cite{schrijver2007} and \cite{welsch+li2008} infer that,
statistically speaking, the most common driver of the formation of the
SHIL is flux emergence. That is confirmed by the present sample,
although AR\,11158 and possibly AR\,11166 appear to be examples in
which the SHILs form by shearing coalescence of previously emerged
structures.

SHILs are transient, surviving typically for 1-2\,days. This was noted
in studies of relatively small samples
\cite[including][]{pevtsov+etal1994,schrijver+etal2005a,2014ApJ...797...50A}.
\cite{2010ApJ...723..634M} performed a large-sample analysis of
magnetograms of flaring regions using {\em SOHO/MDI} magnetograms, reviewing
data for 1075 active regions and for 6000 C, M, and X flares.  They
find that ``[f]lare occurrence is statistically associated with
changes in several characteristics of the line-of-sight magnetic field
in solar active regions (ARs). \ldots\ The best-calculated parameter,
the gradient-weighted inversion-line length (GWILL), combines the
primary polarity inversion line (PIL) length and the gradient across
it. Therefore, GWILL is sensitive to complex field structures via the
length of the PIL and shearing via the gradient. GWILL shows an
average 35\%\ increase during the 40\,hr prior to X-class flares, a
16\%\ increase before M-class flares, and 17\%\ increase prior to
B-C-class flares,'' followed by a decrease after the flare when
analyzing behavior in a superposed epoch analysis.  The present sample
of regions shows consistent average trends (see
Sect.~\ref{sec:analysis} and Fig.~\ref{fig:logrt}).
Based only on that, one might conclude in a statistical sense that the
flux emergence at the SHIL completes within about a day of the flare
time. The range in behavior for individual regions is so
large, however, that this tendency for flux in the region and the SHIL to
peak around the time of major flaring is not useful as a predictive
indicator of either major flaring or termination of flux emergence for
individual regions.

A case of particular interest here is that of AR\,11944:
whereas this region has a high value of $\log(R)=4.7$ that frequently
accompanies X-class flaring, the flare in fact stems from a filament
eruption in a PIL that has no high-gradient field in it (but with the
filament-carrying flux rope an obvious candidate contributor to the
nonpotential appearance of the coronal field).

It is interesting in this context to note that there are very few
known cases of flares of X magnitude
that occur without involving a SHIL, or in regions in which no SHIL
exists at al. \cite{1980IAUS...91..207M} and
\cite{1982SoPh...78..271S} describe a filament eruption
from an old spotless region. \cite{1980IAUS...91..207M} show a BBSO
videomagnetogram suggesting no SHIL, while the H$\alpha$ filtergrams
\citep[e.g.][]{1979SoPh...64..165M} show a pronounced filament
configuration running along the north-south oriented PIL. As this
event occurred prior to the GOES satellites, whether or not it was
indeed an X-class flare needs to be established indirectly;
\cite{1982SoPh...78..271S} discuss the X-ray observations of this
event, yielding an emission measure of some $10^{50}$\,cm$^{-3}$ and a
temperature of $\approx 8.5$\,MK at peak emission measure (after
cooling from a peak value of 10.4\,MK), not atypical for an X flares.

What is the origin of largely un-neutralized currents that may lead to
AR-scale nonpotentiality?  Modeling work by \cite{2015ApJ...810...17D}
and others cited in Sect.~\ref{sec:introduction} suggests that this
requires shearing of the neutral line, such as is often seen
accompanying flux emergence.  The emergence of flux into pre-existing
configurations can result in high-$R$ SHILs. In the case of intense
flaring, this occurs most frequently when the emerging flux bundle
appears to be 
a current-carrying rope (which typically takes $1-2$\,d to fully
emerge), sometimes when emerging concentrations are pushed into and
along pre-existing clusters of substantial flux (on time
scales of hours to days), while rarely the currents survive or grow as
the active region evolves and SHIL signatures vanish or even until
spots have dissolved (which can take over a week).
 
Current systems leading to nonpotential configurations can also occur
by rotation of flux concentrations, e.g., rotating sunspots (which
would cause swirls in the loops above these, as is
seen in at least 18 of the 37 substantively nonpotential regions in the
sample). The work by \cite{2015ApJ...810...17D} demonstrates that if
such rotating motions are well away from the PIL, then direct and
return currents should neutralize; yet, when these rotating motions
involve much of the flux, the corona is likely still distorted
relative to a potential configuration even if the current should
neutralize overall.

Such bulk rotation of large flux clusters might be caused
by subsurface vortices. Searching for relationships
between surface field and subsurface flows, however,
\cite{2014ApJ...795..113S} ``find no significant region-by-region
correlation between the subsurface kinetic helicity and either the
strong-field current helicity or $\alpha$'' (the ratio of current
density and field strength in a nonlinear force-free field). In fact,
``subsurface fluid motions with a given sign of kinetic
helicity appear to correspond to photospheric field structures of the
same and of opposite handedness in approximately equal numbers.''

Signatures of electrical currents and of shearing (sometimes
rotational) flows are seen as important for flaring behavior of active
regions. \cite{2015ApJ...798..135B}, for example, analyze 1.5 million
observations of 2071 active regions in search for predictors of
flaring activity. They identify four parameters as key discriminants
for flaring: the total unsigned current helicity, total magnitude of
the Lorentz force, total photospheric magnetic free energy density,
and the total unsigned vertical current. All of these metrics can be
derived only from vector-magnetic data. However, there are
correlations that may point to comparably informative quantities that
do not involve vector data. For example, \cite{2014ApJ...788..150S},
based on 3226 vector magnetograms of 61 active regions, show that
total vertical absolute photospheric current and total unsigned flux
in active regions are proportional over two orders of magnitude. Their
sample includes regions that show no flaring of class C or above up to
regions with X-class flaring. The Spearman rank correlation
coefficient of that relationship is 0.98 and spread about the mean
relationship is no wider than about a factor of 1.5 along the full range
in fluxes (their Fig.\,7b).

Net currents can apparently exist in some active regions (and in
smaller spotless bipolar regions) well past the flux-emergence phase;
perhaps, they can even be strengthened by sustained shear flows or by
changes induced by the evolution of neighboring field, such as the
emergence of an adjacent active region. Hence, old, decaying regions,
some even past their spot-sporting phase (if ever they had spots), can
appear nonpotential, often with filament eruptions from within them
of from their interfaces with other (emerging, mature, or decayed)
regions in the vicinity. This inference about sustained currents is to
replace the hypothesis by \cite{schrijver+etal2005a} who estimated
the characteristic life time of current systems large enough to make
the coronal field appear nonpotential to be of order 20-30\,hrs upon
termination of flux emergence, but did so based on a strongly biased
selection of {\em TRACE} target regions.

The observations show all X-class flares to emanate from regions
harboring SHILs, although 3 of 10 of these regions are not obviously
nonpotential in a substantial part of their coronal appearance, and in
one case the flare was not obviously connected to the SHIL. On the
other hand, large regions can appear nonpotential without exhibiting
major flaring. I conclude that the dynamics of the flux involved in
the compact SHILs (which sets the regions' $R$ values) is of
preeminent importance for the large-flare potential of active regions,
but that their corresponding currents may not reveal themselves in
region-scale nonpotentiality, although their distorting field
signatures are often seen low over the SHILs. The sample of X-flaring
regions available to us, admittedly small, is consistent with MHD
studies that suggest that their flux emergence off the main PIL may
shield the effects of the currents involved through neutralizing
counter-currents nearby. In contrast to the SHILs as diagnostic for
flare potential, active region-scale nonpotentiality may inform us
about the eruption potential other than those from SHILs, which almost
never leads to X-class flaring.

\begin{table}[t]
  \caption{Contingency table for nonpotentiality and $\ge$X1 flaring
    in active regions between 2010/05/01 and
    2014/10/31 estimated to meet the selection criteria for this
    study (see Sect.~\ref{sec:analysis}). The $\chi_\nu^2$ test for independence of flaring
    and nonpotentiality yields $\chi_\nu^2=115$.}\label{tab:pfcontingency}
\begin{center}
\begin{tabular}{l|cc|r}
\hline\hline 
No.\ of regions:& near-potential & nonpotential  & total \\
\hline
no flares $\le$M9    & $\approx 950$ &  26 & $\approx 976$ \\
flaring at X1 or above &  3 &  6 & 9 \\
\hline
total & $\approx 953$ & 32 & $\approx 985$\\
\hline
\end{tabular}
\end{center}
\end{table}
 The observations suggest that the electrical currents attending the
nonpotential appearance of active-region coronae and those
accompanying major flaring are significantly correlated
(see the contingency Table~\ref{tab:pfcontingency}), but not necessarily the
same. Intense flaring generally occurs when flux is
emerging, particularly when there are signatures of strong shearing
flow and twisted field. Coronal nonpotentiality may result from
sub-surface motions before the region emerges,
as well as reflect ropes that intrude into the region later
on, thus compounding successive generations of electrical
currents through emergence and induction. Nonpotentiality generally
occurs only in regions in which flux emergence is sufficiently large
relative to existing flux (some 15\%\ or more) and with strong-field
high-gradient polarity inversion lines that are relatively long
compared to the region's size scale and straddle the region's primary
polarity inversion line. Nonpotentiality outlasts the flux-emergence
and flaring phase of the regions' evolution by typically a day, but
infrequently (in about 1 in 200 regions) it can last for over the four
days of the sampling window used here.

\acknowledgements I thank Mark Cheung and Alan Title for their
comments, and Hugh Hudson for pointing out the ``spotless'' X flare of
1973. This work was supported by NASA's {\it {\em SDO}/{\em AIA}}
contract (NNG04EA00C) to LMATC. {\em AIA} is an instrument onboard the
{\em Solar Dynamics Observatory}, a mission for NASA's Living With a
Star program.

%\bibliographystyle{/Users/schryver/tex/apj}
%\bibliography{ref_karel.bib,nonpotentialIrefs.bib}

\begin{thebibliography}{0}%
\makeatletter
\providecommand \@ifxundefined [1]{%
 \@ifx{#1\undefined}
}%
\providecommand \@ifnum [1]{%
 \ifnum #1\expandafter \@firstoftwo
 \else \expandafter \@secondoftwo
 \fi
}%
\providecommand \@ifx [1]{%
 \ifx #1\expandafter \@firstoftwo
 \else \expandafter \@secondoftwo
 \fi
}%
\providecommand \natexlab [1]{#1}%
\providecommand \enquote  [1]{``#1''}%
\providecommand \bibnamefont  [1]{#1}%
\providecommand \bibfnamefont [1]{#1}%
\providecommand \citenamefont [1]{#1}%
\providecommand \href@noop [0]{\@secondoftwo}%
\providecommand \href [0]{\begingroup \@sanitize@url \@href}%
\providecommand \@href[1]{\@@startlink{#1}\@@href}%
\providecommand \@@href[1]{\endgroup#1\@@endlink}%
\providecommand \@sanitize@url [0]{\catcode `\\12\catcode `\$12\catcode
  `\&12\catcode `\#12\catcode `\^12\catcode `\_12\catcode `\%12\relax}%
\providecommand \@@startlink[1]{}%
\providecommand \@@endlink[0]{}%
\providecommand \url  [0]{\begingroup\@sanitize@url \@url }%
\providecommand \@url [1]{\endgroup\@href {#1}{\urlprefix }}%
\providecommand \urlprefix  [0]{URL }%
\providecommand \Eprint [0]{\href }%
\providecommand \doibase [0]{http://dx.doi.org/}%
\providecommand \selectlanguage [0]{\@gobble}%
\providecommand \bibinfo  [0]{\@secondoftwo}%
\providecommand \bibfield  [0]{\@secondoftwo}%
\providecommand \translation [1]{[#1]}%
\providecommand \BibitemOpen [0]{}%
\providecommand \bibitemStop [0]{}%
\providecommand \bibitemNoStop [0]{.\EOS\space}%
\providecommand \EOS [0]{\spacefactor3000\relax}%
\providecommand \BibitemShut  [1]{\csname bibitem#1\endcsname}%
\let\auto@bib@innerbib\@empty
%</preamble>
\end{thebibliography}%


\begin{thebibliography}{}

\bibitem[\protect\citeauthoryear{{Aschwanden} {\em
  et~al.\/}}{2014}]{2014ApJ...797...50A}
{Aschwanden}, M.~J., {Xu}, Y., {\&} {Jing}, J. 2014,
\newblock ApJ. 797, 50

\bibitem[\protect\citeauthoryear{Barnes and Leka}{2008}]{barnes+leka2008}
Barnes, G. {\&} Leka, K.~D. 2008,
\newblock ApJL 688, 107

\bibitem[\protect\citeauthoryear{{Bobra} and
  {Couvidat}}{2015}]{2015ApJ...798..135B}
{Bobra}, M.~G. {\&} {Couvidat}, S. 2015,
\newblock ApJ. 798, 135

\bibitem[\protect\citeauthoryear{{Chen} {\em
  et~al.\/}}{2015}]{2015ApJ...808L..24C}
{Chen}, H., {Zhang}, J., {Ma}, S., {Yang}, S., {Li}, L., {Huang}, X., {\&}
  {Xiao}, J. 2015,
\newblock ApJL 808, L24

\bibitem[\protect\citeauthoryear{{Chifu} {\em
  et~al.\/}}{2015}]{2015AnA...577A.123C}
{Chifu}, I., {Inhester}, B., {\&} {Wiegelmann}, T. 2015,
\newblock A{\&}A 577, A123

\bibitem[\protect\citeauthoryear{{Chintzoglou} {\em
  et~al.\/}}{2015}]{2015ApJ...809...34C}
{Chintzoglou}, G., {Patsourakos}, S., {\&} {Vourlidas}, A. 2015,
\newblock ApJ. 809, 34

\bibitem[\protect\citeauthoryear{{Chintzoglou} and
  {Zhang}}{2013}]{2013ApJ...764L...3C}
{Chintzoglou}, G. {\&} {Zhang}, J. 2013,
\newblock ApJL 764, L3

\bibitem[\protect\citeauthoryear{{Dalmasse} {\em
  et~al.\/}}{2015}]{2015ApJ...810...17D}
{Dalmasse}, K., {Aulanier}, G., {D{\'e}moulin}, P., {Kliem}, B.,
  {T{\"o}r{\"o}k}, T., {\&} {Pariat}, E. 2015,
\newblock ApJ. 810, 17

\bibitem[\protect\citeauthoryear{{DeRosa} {\em
  et~al.\/}}{2015}]{2015arXiv150805455D}
{DeRosa}, M.~L., {Wheatland}, M.~S., {Leka}, K.~D., {Barnes}, G., {Amari}, T.,
  {Canou}, A., {Gilchrist}, S.~A., {Thalmann}, J.~K., {Valori}, G.,
  {Wiegelmann}, T., {Schrijver}, C.~J., {Malanushenko}, A., {Sun}, X., {\&}
  {R{\'e}gnier}, S. 2015,
\newblock ApJ. 811, 107

\bibitem[\protect\citeauthoryear{{Fang} {\em
  et~al.\/}}{2012}]{2012ApJ...754...15F}
{Fang}, F., {Manchester}, IV, W., {Abbett}, W.~P., {\&} {van der Holst}, B.
  2012,
\newblock ApJ. 754, 15

\bibitem[\protect\citeauthoryear{Handy {\em et~al.\/}}{1999}]{traceinstrument}
Handy, B.~N., Acton, L.~W., Kankelborg, C.~C., Wolfson, C.~J., Akin, D.~J.,
  Bruner, M.~E., Carvalho, R., Catura, R.~C., Chevalier, R., Duncan, D.~W.,
  Edwards, C.~G., Feinstein, C.~N., Freeland, S.~L., Friedlander, F.~M.,
  Hoffman, C.~H., Hurlburt, N.~E., Jurcevich, B.~K., Katz, N.~L., Kelly, G.~A.,
  Lemen, J.~R., Levay, M., Lindgren, R.~W., Mathur, D.~P., Meyer, S.~B.,
  Morrison, S.~J., Morrison, M.~D., Nightingale, R.~W., Pope, T.~P., Rehse,
  R.~A., Schrijver, C.~J., Shine, R.~A., Shing, L., Strong, K.~T., Tarbell,
  T.~D., Title, A.~M., Torgerson, D.~D., Golub, L., Bookbinder, J.~A.,
  Caldwell, D., Cheimets, P.~N., Davis, W.~N., Deluca, E.~E., McMullen, R.~A.,
  Amato, D., Fisher, R., Maldonado, H., {\&} Parkinson, C. 1999,
\newblock Solar Phys. 187, 229

\bibitem[\protect\citeauthoryear{{Inoue} {\em
  et~al.\/}}{2015}]{2015ApJ...803...73I}
{Inoue}, S., {Hayashi}, K., {Magara}, T., {Choe}, G.~S., {\&} {Park}, Y.~D.
  2015,
\newblock ApJ. 803, 73

\bibitem[\protect\citeauthoryear{{Janvier} {\em
  et~al.\/}}{2014}]{2014ApJ...788...60J}
{Janvier}, M., {Aulanier}, G., {Bommier}, V., {Schmieder}, B., {D{\'e}moulin},
  P., {\&} {Pariat}, E. 2014,
\newblock ApJ. 788, 60

\bibitem[\protect\citeauthoryear{{Kazachenko} {\em
  et~al.\/}}{2015}]{2015ApJ...811...16K}
{Kazachenko}, M.~D., {Fisher}, G.~H., {Welsch}, B.~T., {Liu}, Y., {\&} {Sun},
  X. 2015,
\newblock ApJ. 811, 16

\bibitem[\protect\citeauthoryear{{Komm} {\em
  et~al.\/}}{2014}]{2014SoPh..289..475K}
{Komm}, R., {Gosain}, S., {\&} {Pevtsov}, A. 2014,
\newblock Solar Phys. 289, 475

\bibitem[\protect\citeauthoryear{Leibacher {\em
  et~al.\/}}{2010}]{2010SoPh..263....1.}
Leibacher, J., Sakurai, T., Schrijver, C.~J., {\&} {van Driel-Gesztelyi}, L.
  2010,
\newblock Solar Phys. 263, 1

\bibitem[\protect\citeauthoryear{{Leka} and {Barnes}}{2007}]{leka+barnes2007}
{Leka}, K.~D. {\&} {Barnes}, G. 2007,
\newblock ApJ. 656, 1173

\bibitem[\protect\citeauthoryear{{Lemen} {\em et~al.\/}}{2012}]{aiainstrument}
{Lemen}, J.~R., {Title}, A.~M., {Akin}, D.~J., {Boerner}, P.~F., {Chou}, C.,
  {Drake}, J.~F., {Duncan}, D.~W., {Edwards}, C.~G., {Friedlaender}, F.~M.,
  {Heyman}, G.~F., {Hurlburt}, N.~E., {Katz}, N.~L., {Kushner}, G.~D., {Levay},
  M., {Lindgren}, R.~W., {Mathur}, D.~P., {McFeaters}, E.~L., {Mitchell}, S.,
  {Rehse}, R.~A., {Schrijver}, C.~J., {Springer}, L.~A., {Stern}, R.~A.,
  {Tarbell}, T.~D., {Wuelser}, J.-P., {Wolfson}, C.~J., {Yanari}, C.,
  {Bookbinder}, J.~A., {Cheimets}, P.~N., {Caldwell}, D., {Deluca}, E.~E.,
  {Gates}, R., {Golub}, L., {Park}, S., {Podgorski}, W.~A., {Bush}, R.~I.,
  {Scherrer}, P.~H., {Gummin}, M.~A., {Smith}, P., {Auker}, G., {Jerram}, P.,
  {Pool}, P., {Soufli}, R., {Windt}, D.~L., {Beardsley}, S., {Clapp}, M.,
  {Lang}, J., {\&} {Waltham}, N. 2012,
\newblock Solar Phys. 275, 17

\bibitem[\protect\citeauthoryear{{Li} and {Liu}}{2015}]{2015SoPh..290.2199L}
{Li}, A. {\&} {Liu}, Y. 2015,
\newblock Solar Phys. 290, 2199

\bibitem[\protect\citeauthoryear{{Liu} {\em
  et~al.\/}}{2013}]{2013SoPh..287..279L}
{Liu}, Y., {Zhao}, J., {\&} {Schuck}, P.~W. 2013,
\newblock Solar Phys. 287, 279

\bibitem[\protect\citeauthoryear{{Malanushenko} {\em
  et~al.\/}}{2014}]{2014ApJ...783..102M}
{Malanushenko}, A., {Schrijver}, C.~J., {DeRosa}, M.~L., {\&} {Wheatland},
  M.~S. 2014,
\newblock ApJ. 783, 102

\bibitem[\protect\citeauthoryear{{Malanushenko} {\em
  et~al.\/}}{2012}]{2012ApJ...756..153M}
{Malanushenko}, A., {Schrijver}, C.~J., {DeRosa}, M.~L., {Wheatland}, M.~S.,
  {\&} {Gilchrist}, S.~A. 2012,
\newblock ApJ. 756, 153

\bibitem[\protect\citeauthoryear{{Manchester} {\em
  et~al.\/}}{2004}]{manchester+etal2004}
{Manchester}, W., {Gombosi}, T., {DeZeeuw}, D., {\&} {Fan}, Y. 2004,
\newblock ApJ. 610, 588

\bibitem[\protect\citeauthoryear{{Martin}}{1979}]{1979SoPh...64..165M}
{Martin}, S.~F. 1979,
\newblock Solar Phys. 64, 165

\bibitem[\protect\citeauthoryear{{Mason} and
  {Hoeksema}}{2010}]{2010ApJ...723..634M}
{Mason}, J.~P. {\&} {Hoeksema}, J.~T. 2010,
\newblock ApJ. 723, 634

\bibitem[\protect\citeauthoryear{{Metcalf} {\em
  et~al.\/}}{2008}]{metcalf+etal2007}
{Metcalf}, T.~R., {Derosa}, M.~L., {Schrijver}, C.~J., {Barnes}, G., {van
  Ballegooijen}, A.~A., {Wiegelmann}, T., {Wheatland}, M.~S., {Valori}, G.,
  {\&} {McTtiernan}, J.~M. 2008,
\newblock Solar Phys. 247, 269

\bibitem[\protect\citeauthoryear{{Moore} and
  {Labonte}}{1980}]{1980IAUS...91..207M}
{Moore}, R.~L. {\&} {Labonte}, B.~J. 1980,
\newblock in M. {Dryer} and E. {Tandberg-Hanssen} (Eds.), {\em Solar and
  Interplanetary Dynamics\/}, Vol.~91 of {\em IAU Symposium\/}, p.~207

\bibitem[\protect\citeauthoryear{{Pesnell} {\em
  et~al.\/}}{2012}]{2012SoPh..275....3P}
{Pesnell}, W.~D., {Thompson}, B.~J., {\&} {Chamberlin}, P.~C. 2012,
\newblock Solar Phys. 275, 3

\bibitem[\protect\citeauthoryear{{Petrie}}{2012}]{2012ApJ...759...50P}
{Petrie}, G.~J.~D. 2012,
\newblock ApJ. 759, 50

\bibitem[\protect\citeauthoryear{{Petrie}}{2013}]{2013SoPh..287..415P}
{Petrie}, G.~J.~D. 2013,
\newblock Solar Phys. 287, 415

\bibitem[\protect\citeauthoryear{{Pevtsov} {\em
  et~al.\/}}{1994}]{pevtsov+etal1994}
{Pevtsov}, A.~A., {Canfield}, R.~C., {\&} {Metcalf}, T.~R. 1994,
\newblock ApJL 425, 117

\bibitem[\protect\citeauthoryear{{Ruan} {\em
  et~al.\/}}{2015}]{2015ApJ...812..120R}
{Ruan}, G., {Chen}, Y., {\&} {Wang}, H. 2015,
\newblock ApJ. 812, 120

\bibitem[\protect\citeauthoryear{{Ruan} {\em
  et~al.\/}}{2014}]{2014ApJ...784..165R}
{Ruan}, G., {Chen}, Y., {Wang}, S., {Zhang}, H., {Li}, G., {Jing}, J., {Su},
  J., {Li}, X., {Xu}, H., {Du}, G., {\&} {Wang}, H. 2014,
\newblock ApJ. 784, 165

\bibitem[\protect\citeauthoryear{{Savcheva} {\em
  et~al.\/}}{2014}]{2014SoPh..289.3297S}
{Savcheva}, A.~S., {McKillop}, S.~C., {McCauley}, P.~I., {Hanson}, E.~M., {\&}
  {DeLuca}, E.~E. 2014,
\newblock Solar Phys. 289, 3297

\bibitem[\protect\citeauthoryear{Scherrer {\em et~al.\/}}{1995}]{soho}
Scherrer, P.~H., Bogart, R.~S., Bush, R.~I., Hoeksema, J.~T., Kosovichev,
  A.~G., Schou, J., Rosenberg, W., Springer, L., Tarbell, T.~D., Title, A.,
  Wolfson, C.~J., Zayer, I., {\&} {The MDI Engineering Team} 1995,
\newblock Solar Phys. 162, 129

\bibitem[\protect\citeauthoryear{{Scherrer} {\em
  et~al.\/}}{2012}]{2012SoPh..275..207S}
{Scherrer}, P.~H., {Schou}, J., {Bush}, R.~I., {Kosovichev}, A.~G., {Bogart},
  R.~S., {Hoeksema}, J.~T., {Liu}, Y., {Duvall}, T.~L., {Zhao}, J., {Title},
  A.~M., {Schrijver}, C.~J., {Tarbell}, T.~D., {\&} {Tomczyk}, S. 2012,
\newblock Solar Phys. 275, 207

\bibitem[\protect\citeauthoryear{{Schrijver}}{2007}]{schrijver2007}
{Schrijver}, C.~J. 2007,
\newblock ApJL 655, 117

\bibitem[\protect\citeauthoryear{{Schrijver}}{2009}]{schrijver2009b}
{Schrijver}, C.~J. 2009,
\newblock Advances in Space Research 43, 739

\bibitem[\protect\citeauthoryear{{Schrijver} {\em
  et~al.\/}}{2011}]{2011ApJ...738..167S}
{Schrijver}, C.~J., {Aulanier}, G., {Title}, A.~M., {Pariat}, E., {\&}
  {Delann{\'e}e}, C. 2011,
\newblock ApJ. 738, 167

\bibitem[\protect\citeauthoryear{{Schrijver} {\em
  et~al.\/}}{2005}]{schrijver+etal2005a}
{Schrijver}, C.~J., {De Rosa}, M.~L., Title, A.M., {\&} {Metcalf}, T.~R. 2005,
\newblock ApJ. 628, 501

\bibitem[\protect\citeauthoryear{Schrijver and
  DeRosa}{2003}]{schrijver+derosa2002b}
Schrijver, C.~J. {\&} DeRosa, M.~L. 2003,
\newblock Solar Phys. 212, 165

\bibitem[\protect\citeauthoryear{{Schrijver} and
  {Title}}{2011}]{schrijver+title2010}
{Schrijver}, C.~J. {\&} {Title}, A.~M. 2011,
\newblock Journal of Geophysical Research (Space Physics) 116(A15), 4108

\bibitem[\protect\citeauthoryear{{Schrijver} {\em
  et~al.\/}}{2013}]{2013ApJ...773...93S}
{Schrijver}, C.~J., {Title}, A.~M., {Yeates}, A.~R., {\&} {DeRosa}, M.~L. 2013,
\newblock ApJ. 773, 93

\bibitem[\protect\citeauthoryear{{Seligman} {\em
  et~al.\/}}{2014}]{2014ApJ...795..113S}
{Seligman}, D., {Petrie}, G.~J.~D., {\&} {Komm}, R. 2014,
\newblock ApJ. 795, 113

\bibitem[\protect\citeauthoryear{{Su} {\em
  et~al.\/}}{2014}]{2014ApJ...788..150S}
{Su}, J.~T., {Jing}, J., {Wang}, S., {Wiegelmann}, T., {\&} {Wang}, H.~M. 2014,
\newblock ApJ. 788, 150

\bibitem[\protect\citeauthoryear{{Sun} {\em
  et~al.\/}}{2015}]{2015ApJ...804L..28S}
{Sun}, X., {Bobra}, M.~G., {Hoeksema}, J.~T., {Liu}, Y., {Li}, Y., {Shen}, C.,
  {Couvidat}, S., {Norton}, A.~A., {\&} {Fisher}, G.~H. 2015,
\newblock ApJL 804, L28

\bibitem[\protect\citeauthoryear{{Sun} {\em
  et~al.\/}}{2012}]{2012ApJ...748...77S}
{Sun}, X., {Hoeksema}, J.~T., {Liu}, Y., {Wiegelmann}, T., {Hayashi}, K.,
  {Chen}, Q., {\&} {Thalmann}, J. 2012,
\newblock ApJ. 748, 77

\bibitem[\protect\citeauthoryear{{Svestka} {\em
  et~al.\/}}{1982}]{1982SoPh...78..271S}
{Svestka}, Z., {Dodson-Prince}, H.~W., {Mohler}, O.~C., {Martin}, S.~F.,
  {Moore}, R.~L., {Nolte}, J.~T., {\&} {Petrasso}, R.~D. 1982,
\newblock Solar Phys. 78, 271

\bibitem[\protect\citeauthoryear{{Toriumi} {\em
  et~al.\/}}{2014}]{2014SoPh..289.3351T}
{Toriumi}, S., {Iida}, Y., {Kusano}, K., {Bamba}, Y., {\&} {Imada}, S. 2014,
\newblock Solar Phys. 289, 3351

\bibitem[\protect\citeauthoryear{{T{\"o}r{\"o}k} and
  {Kliem}}{2003}]{torok+kliem2003}
{T{\"o}r{\"o}k}, T. {\&} {Kliem}, B. 2003,
\newblock A{\&}A 406, 1043

\bibitem[\protect\citeauthoryear{{T{\"o}r{\"o}k} {\em
  et~al.\/}}{2014}]{2014ApJ...782L..10T}
{T{\"o}r{\"o}k}, T., {Leake}, J.~E., {Titov}, V.~S., {Archontis}, V.,
  {Miki{\'c}}, Z., {Linton}, M.~G., {Dalmasse}, K., {Aulanier}, G., {\&}
  {Kliem}, B. 2014,
\newblock ApJL 782, L10

\bibitem[\protect\citeauthoryear{{Vemareddy} {\em
  et~al.\/}}{2012}]{2012ApJ...761...86V}
{Vemareddy}, P., {Ambastha}, A., {Maurya}, R.~A., {\&} {Chae}, J. 2012,
\newblock ApJ. 761, 86

\bibitem[\protect\citeauthoryear{{Vemareddy} and
  {Wiegelmann}}{2014}]{2014ApJ...792...40V}
{Vemareddy}, P. {\&} {Wiegelmann}, T. 2014,
\newblock ApJ. 792, 40

\bibitem[\protect\citeauthoryear{{Wang} and {Liu}}{2015}]{2015RAA....15..145W}
{Wang}, H. {\&} {Liu}, C. 2015,
\newblock Research in Astronomy and Astrophysics 15, 145

\bibitem[\protect\citeauthoryear{{Wang} {\em
  et~al.\/}}{2014}]{2014ApJ...781L..23W}
{Wang}, H., {Liu}, C., {Deng}, N., {Zeng}, Z., {Xu}, Y., {Jing}, J., {\&}
  {Cao}, W. 2014,
\newblock ApJL 781, L23

\bibitem[\protect\citeauthoryear{{Wang} {\em
  et~al.\/}}{2013}]{2013SoPh..288..507W}
{Wang}, R., {Yan}, Y., {\&} {Tan}, B. 2013,
\newblock Solar Phys. 288, 507

\bibitem[\protect\citeauthoryear{{Welsch} and {Li}}{2008}]{welsch+li2008}
{Welsch}, B.~T. {\&} {Li}, Y. 2008,
\newblock in R. {Howe}, R.~W. {Komm}, K.~S. {Balasubramaniam}, and G.~J.~D.
  {Petrie} (Eds.), {\em Subsurface and Atmospheric Influences on Solar
  Activity\/}, Vol. 383 of {\em Astron.\ Soc.\ Pacific CS-\/},  429

\end{thebibliography}

\end{document}